\documentclass[aps,prc,twocolumn, superscriptaddress,showkeys,nofootinbib]{revtex4-1}  

\usepackage[utf8]{inputenc}

\usepackage{bibentry}

\usepackage{graphicx,color}
\usepackage{amsmath,amssymb,amsfonts}
\usepackage{hyperref}

\usepackage{xspace}	

\definecolor{UTOrange}{rgb}{1, 0.51, 0.0}

\begin{document}
\title{
Model Comparisons of Transverse Energy and Charged-Particle Multiplicity in A+A Collisions at Midrapidity from $\sqrt{s_{NN}}$ $=$ 7.7 to 200~GeV
}
\medskip
\author{Niseem~Magdy} 
\email{niseem.abdelrahman@tsu.edu}
\affiliation{Department of Physics, University of Tennessee, Knoxville, TN, 37996, USA}
\affiliation{Department of Physics, Texas Southern University, Houston, TX 77004, USA}
\affiliation{Physics Department, Brookhaven National Laboratory, Upton, New York 11973, USA}
\author{Antonio~Silva}  
\affiliation{Department of Physics, University of Tennessee, Knoxville, TN, 37996, USA}
\affiliation{Department of Physics and Astronomy, Iowa State University, Iowa, 50011, USA}
\author{Christal~Martin} 
\affiliation{Department of Physics, University of Tennessee, Knoxville, TN, 37996, USA}
\author{Josie~Hakanson} 
\affiliation{Department of Physics, University of Tennessee, Knoxville, TN, 37996, USA}
\author{Olivia~Bartoshesky} 
\affiliation{Department of Physics, University of Tennessee, Knoxville, TN, 37996, USA}
\author{Aidan~Hill} 
\affiliation{Department of Physics, University of Tennessee, Knoxville, TN, 37996, USA}
\author{Christine Nattrass}
\email{cnattras@utk.edu}
\affiliation{Department of Physics, University of Tennessee, Knoxville, TN, 37996, USA}




\begin{abstract}
We present a comprehensive comparison of PHENIX measurements of transverse energy production ($dE_T/d\eta$) and charged-particle multiplicity ($dN/d\eta$) at midrapidity to simulations from PYTHIA-8, AMPT, HIJING, and SMASH. These comparisons span both small systems (d+Au, $^3$He+Au) and large systems (Cu+Cu, Cu+Au, Au+Au, and U+U) at $\sqrt{s_{NN}}$ $\sim$ 200~GeV and Au+Au over a range of beam energies $\sqrt{s_{NN}} = 7.7$--200~GeV. Using the Rivet framework, we assess the performance of these models. While general trends are captured, significant deviations persist, particularly in low-energy and peripheral collisions, underscoring the need for improved modeling of baryon stopping and energy deposition mechanisms.
\end{abstract}

\maketitle

\section{Introduction}
Relativistic heavy-ion collisions allow the study of matter at high energy density and temperature. At these extreme conditions, lattice Quantum Chromodynamics (QCD) predicts a phase transition of hadronic matter into a state where quarks and gluons are deconfined, the Quark-Gluon Plasma (QGP)~\cite{Heinz:2000bk, Busza:2018rrf, Heinz:2013th}. Investigations at the Relativistic Heavy Ion Collider (RHIC) and the Large Hadron Collider (LHC) have demonstrated a wide array of complex phenomena related to the QGP~\cite{Shuryak:1978ij, Shuryak:1980tp, Muller:2012zq, STAR:2005gfr, PHENIX:2004vcz, BRAHMS:2004adc, PHOBOS:2004zne}. Among the key observables used to probe these phenomena are the transverse energy density and the charged-particle multiplicity, denoted as $dE_T/d\eta$ and $dN/d\eta$, respectively, which provide essential insights into the initial energy and particle production mechanisms per unit of pseudorapidity in high-energy collisions~\cite{PHENIX:2015tbb, PHENIX:2004vdg, Cleymans:2007uk, Bozek:2005eu, Kharzeev:2001yq}.

The system size and beam energy dependence of the $dE_T/d\eta$ and $dN/d\eta$ can constrain energy deposition mechanisms and particle production processes in models across a wide range of collision systems and energies~\cite{PHENIX:2015tbb, PHENIX:2004vdg}. Systematic data-model comparisons of $dE_T/d\eta$ and $dN/d\eta$ provide a detailed mapping of the nuclear geometry and reveal the interplay between soft and hard scattering processes~\cite{Mendenhall:2020fil, Auvinen:2020mpc, Karpenko:2015xea, Jahan:2024wpj}.

Providing data-model comparisons of the system size dependence spanning small systems such as \( d+\text{Au} \) and \({}^3\text{He}+\text{Au}\) to large systems like \(\text{Au}+\text{Au}\) and \(\text{U}+\text{U}\) is crucial for disentangling initial-state effects and the role of geometry in particle production. These comparisons are essential for understanding how the medium evolves across system sizes and identifying potential thresholds for the emergence of collective effects~\cite{PHENIX:2013ktj, PHENIX:2022nht, STAR:2022pfn, STAR:2023wmd, ATLAS:2015hzw, CMS:2012qk, ALICE:2012eyl, CMS:2016fnw,  Schenke:2021mxx}.  
Expanding these comparisons to include multiple beam energies is equally essential, as it enables the exploration of the QCD phase diagram over a broad range of temperatures (T) and baryon chemical potentials (\(\mu_B\))~\cite{Aoki:2006we, Borsanyi:2020fev, Bazavov:2011nk, HotQCD:2018pds}. At high energies, such as \(\sqrt{s_{NN}} = 200 \, \text{GeV}\), the medium probes conditions of high temperature and low \(\mu_B\), where lattice QCD predicts a smooth crossover transition between hadronic matter and quark-gluon plasma~\cite{Guenther:2020jwe, Fodor:2001au, Fodor:2001pe, Csikor:2002ic}. In contrast, lower energies probe regions of higher \(\mu_B\), approaching the domain where models suggest a first-order phase transition and a critical endpoint. Consequently, extending \( dE_T/d\eta \) and \( dN_{\text{ch}}/d\eta \) data-model comparisons over this energy range could shed light on the dynamics of the phase transition and the influence of energy density on medium properties~\cite{Magdy:2021cci, STAR:2012och, Becattini:2005xt, Jahan:2024wpj, Werner:2024ntd, Cimerman:2023hjw, STAR:2022vkx, STAR:2022gki, Rao:2019vgy, STAR:2017ieb}.  

The importance of data and model comparisons lies in their ability to bridge experimental measurements and theoretical predictions, enabling validation and improvement of event generators and other computational models. By allowing direct comparison between experimental measurements and event generators' Monte Carlo predictions, these comparisons ensure consistency and provide a robust testbed for refining theoretical frameworks. The Rivet framework~\cite{Buckley:2010ar} provides a standardized platform for implementing and reproducing analysis routines directly aligned with published data. This capability allows researchers to efficiently compare data to models, contributing significantly to the validation, tuning, and development of theoretical models~\cite{Bierlich:2019rhm}.


The PHENIX collaboration measured $dE_{T}/d\eta$ and $dN/d\eta$ distributions for Au+Au collisions at $\sqrt{s_{NN}} =$ 7.7--200 (GeV), Cu+Cu collisions at $\sqrt{s_{NN}} =$ 200 (GeV) and 62.4 (GeV), U+U collisions at $\sqrt{s_{NN}} =$ 193 (GeV), Cu+Au, $d$+Au, and ${}^3$He+Au collisions at $\sqrt{s_{NN}} =$ 200 (GeV)~\cite{PHENIX:2001kdi, PHENIX:2013ehw, PHENIX:2015tbb}.
We use Rivet to compare these data to PYTHIA-8 Angantyr~\cite{Bierlich:2018xfw}, AMPT~\cite{Lin:2004en}, HIJING~\cite{Wang:1991hta, Gyulassy:1994ew}, and SMASH~\cite{SMASH:2016zqf}.

\section{Methodology}\label{sec:2}
\subsection{Models}
The analysis is performed using events simulated by the PYTHIA-8 Angantyr~\cite{Bierlich:2018xfw}, AMPT (v2.26t9b)~\cite{Lin:2004en}, HIJING (v1.411)~\cite{Wang:1991hta, Gyulassy:1994ew}, and SMASH~\cite{SMASH:2016zqf} models for different energies and system sizes.
This study simulated several data sets summarized in Tab~\ref{tab:1}.  The analysis focuses on minimum bias collisions, with approximately 5.0 million events generated for each case considered.
\begin{table}[h!]
\begin{center}
\caption{The summary of the data sets used in this work.\label{tab:1}}
 \begin{tabular}{|c|c|c|c|}
 \hline 
 System       &  $\sqrt(s_{NN})$  (GeV)\\
 \hline
 Au+Au        &      200, 130, 62.4, 39, 27, 19.6, 14.5, and 7.7 \\   
 \hline
 U+U          &      193                  \\
 \hline
 Cu+Cu        &      200 and 62.4         \\
 \hline
 Cu+Au        &      200                  \\
 \hline
 $^{3}He$+Au  &      200                  \\
 \hline
 $d$+Au       &      200                  \\
 \hline
\end{tabular} 
\end{center}
\end{table}

\subsubsection{PYTHIA-8 Angantyr}
PYTHIA~\cite{Bierlich:2018xfw} is an event generator for studying proton-proton and proton-lepton collisions. In pp collisions, multi-parton interactions (MPI) are generated, assuming that each partonic interaction is mainly independent.  The introduction of the Angantyr model~\cite{Bierlich:2018xfw} extends PYTHIA-8's capabilities to heavy-ion collisions, allowing for the study of $\gamma$-nucleus ($\gamma$A), proton-nucleus (p+A), and nucleus-nucleus (A+A) interactions. Angantyr combines multiple nucleon-nucleon collisions into a single heavy-ion collision. It incorporates various theoretical models suitable for producing hard and soft interactions, initial and final-state parton showers, particle fragmentations, multi-partonic interactions, color reconnection (CR) mechanisms, and decay topologies.

In the current version of the PYTHIA-8 Angantyr model~\cite{Bierlich:2018xfw}, a heavy-ion collision is simulated as a series of nucleon-nucleon interactions, with each projectile nucleon potentially interacting with multiple target nucleons. The number of participating nucleons is determined using the Glauber and Glissando models~\cite{Glauber:1955qq, Rybczynski:2013yba}. Angantyr includes several algorithms to differentiate between various types of nucleon-nucleon interactions, such as elastic, diffractive, and absorptive. This model is designed to accurately describe final-state properties, including multiplicity and transverse momentum distributions, in A+A collisions.
Also, Angantyr does not directly include any mechanism for modeling the QGP medium formed in A+A collisions, although some of the multi-parton interactions may be able to describe the QGP.

\subsubsection{HIJING}
 HIJING~\cite{Wang:1991hta, Gyulassy:1994ew} is a Monte Carlo event generator designed to simulate parton and particle production in high-energy heavy ion collisions. It employs Glauber geometry to simulate heavy-ion collisions via binary nucleon-nucleon interactions. It is used for studying jet and mini-jet production alongside associated particle production in high-energy p+p, p+A, and A+A collisions. The HIJING model integrates PYTHIA~\cite{Sjostrand:1986hx} for determining kinematic variables of scattered partons in each hard- or semihard-interaction, and Lund string fragmentation~\cite{Andersson:1983ia} for the hadronization process.

\subsubsection{AMPT}
AMPT~\cite{Lin:2004en, Ma:2016fve, Ma:2013gga, Ma:2013uqa, Bzdak:2014dia, Nie:2018xog, Haque:2019vgi, Zhao:2019kyk, Bhaduri:2010wi, Nasim:2010hw, Xu:2010du, Magdy:2020bhd, Guo:2019joy, Magdy:2020gxf, Magdy:2022cvt} encompasses several key components: (i) an initial partonic state provided by HIJING~\cite{Wang:1991hta, Gyulassy:1994ew}, with parameters specified in the Lund string fragmentation function~\cite{Ferreres-Sole:2018vgo} denoted by $f(z) \propto z^{-1} (1-z)^a\exp (-b~m_{\perp}^2/z)$, where $a=0.55$ and $b=0.15$ GeV$^{-2}$, as detailed in Ref~\cite{Xu:2011fi}. Here, $z$ represents the light-cone momentum fraction of the produced hadron with transverse mass $m_\perp$ around that of the fragmenting string. (ii) Partonic scattering characterized by a cross-section,
\begin{eqnarray}\label{eq:21}
\sigma_{pp} &=& \dfrac{9 \pi \alpha^{2}_{s}}{2 \mu^{2}}.
\end{eqnarray}
In this context, $\mu$ denotes the screening mass, while $\alpha_{s}$ represents the QCD coupling constant, typically characterizing the expansion dynamics of A+A collisions~\cite{Zhang:1997ej}. Moreover, the model incorporates (iii) the hadronization process through coalescence followed by hadronic interactions~\cite{Li:1995pra}.



\begin{figure*}[t]
 \centering{
 \includegraphics[width=\textwidth,angle=0]{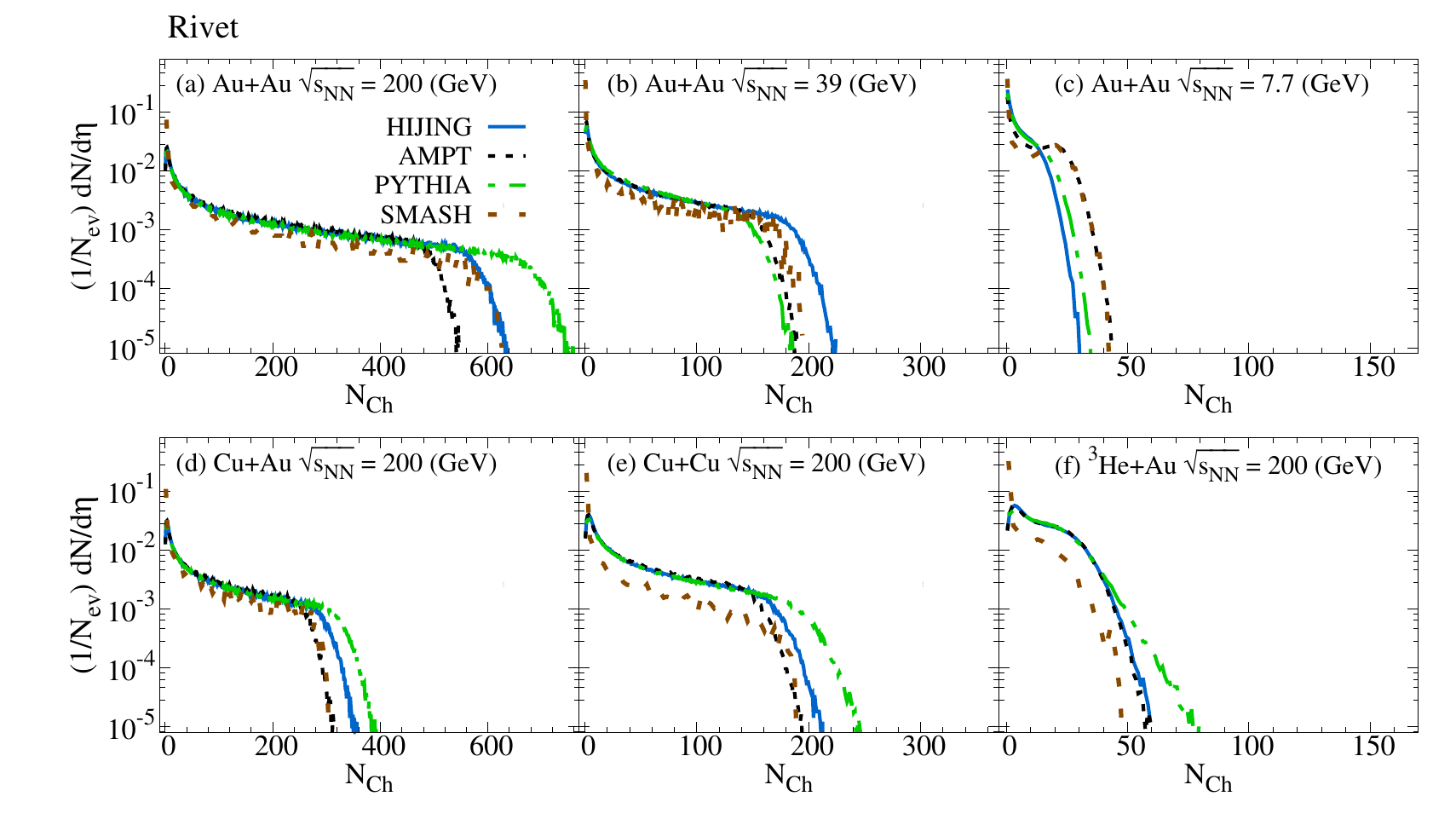}
\vskip -0.36cm
 \caption{
 Comparison of the beam energy and system size dependence of the \(dN/d\eta\) for different MC-models calculations.
 \label{fig:01}
 }
 \vskip -0.46cm
 }
\end{figure*}

\subsubsection{SMASH Transport Model}

SMASH (Simulating Many Accelerated Strongly-interacting Hadrons)~\cite{SMASH:2016zqf} 3.1~\cite{wergieluk_2024_10707746} is a microscopic hadronic transport approach developed to simulate the non-equilibrium dynamics of hadronic matter in heavy-ion collisions, especially at low to intermediate energies where hadronic interactions dominate~\cite{SMASH:2016zqf}. The model solves the relativistic Boltzmann equation for a system of hadrons, treating particles as point-like test particles that propagate and interact via binary collisions, resonance excitations, and decays with vacuum properties. It incorporates a comprehensive set of well-established hadronic states up to masses of about 2~GeV, implementing their spectral functions and decay channels based on experimentally known properties~\cite{SMASH:2016zqf}.

The collision criterion in SMASH follows a geometric interpretation, similar to that used in the UrQMD model~\cite{Bass:1998ca}, where collisions occur when the transverse distance between particles is smaller than a threshold determined by their total cross section. The model offers flexible time evolution algorithms, including fixed time steps or dynamically determined steps based on collision times, and includes optional mean-field potentials to describe nuclear interactions at low energies. SMASH employs a grid-based nearest neighbor search algorithm to efficiently identify collision partners within the system~\cite{SMASH:2016zqf}. Moreover, it can be used as a hadronic afterburner coupled to hydrodynamic simulations to describe the dilute late stages of heavy-ion collisions at higher energies.

\subsection{Analysis}

Rivet is a robust framework designed for comparing Monte Carlo event generators to experimental data from high-energy particle collisions~\cite{Sjostrand:2014zea, Bellm:2015jjp, Hoeche:2012yf, Biro:2019ijx, Krauss:2018djz, Hoche:2018gti, Hoeche:2011fd, Buckley:2016bhy}. 
It has been widely used in the development, validation, and tuning of event generators for Standard Model processes and in studies of parton density functions~\cite{Neill:2018wtk, Reyer:2019obz, Bothmann:2018trh, Proceedings:2018jsb, LHCHiggsCrossSectionWorkingGroup:2016ypw}. The framework supports theoretical and experimental analyses by facilitating comparisons between simulated and experimental results. 
Rivet takes HEPMC~\cite{Dobbs:2001ck, Buckley:2019xhk} output as input and allows a single analysis to be used for every Monte Carlo calculation.
We have implemented Rivet analyses for PHENIX measurements of charged particle multiplicity and transverse energy~\cite{PHENIX:2001kdi, PHENIX:2013ehw, PHENIX:2015tbb}.

\begin{figure*}[t]
 \centering{
 \includegraphics[width=\textwidth,angle=0]{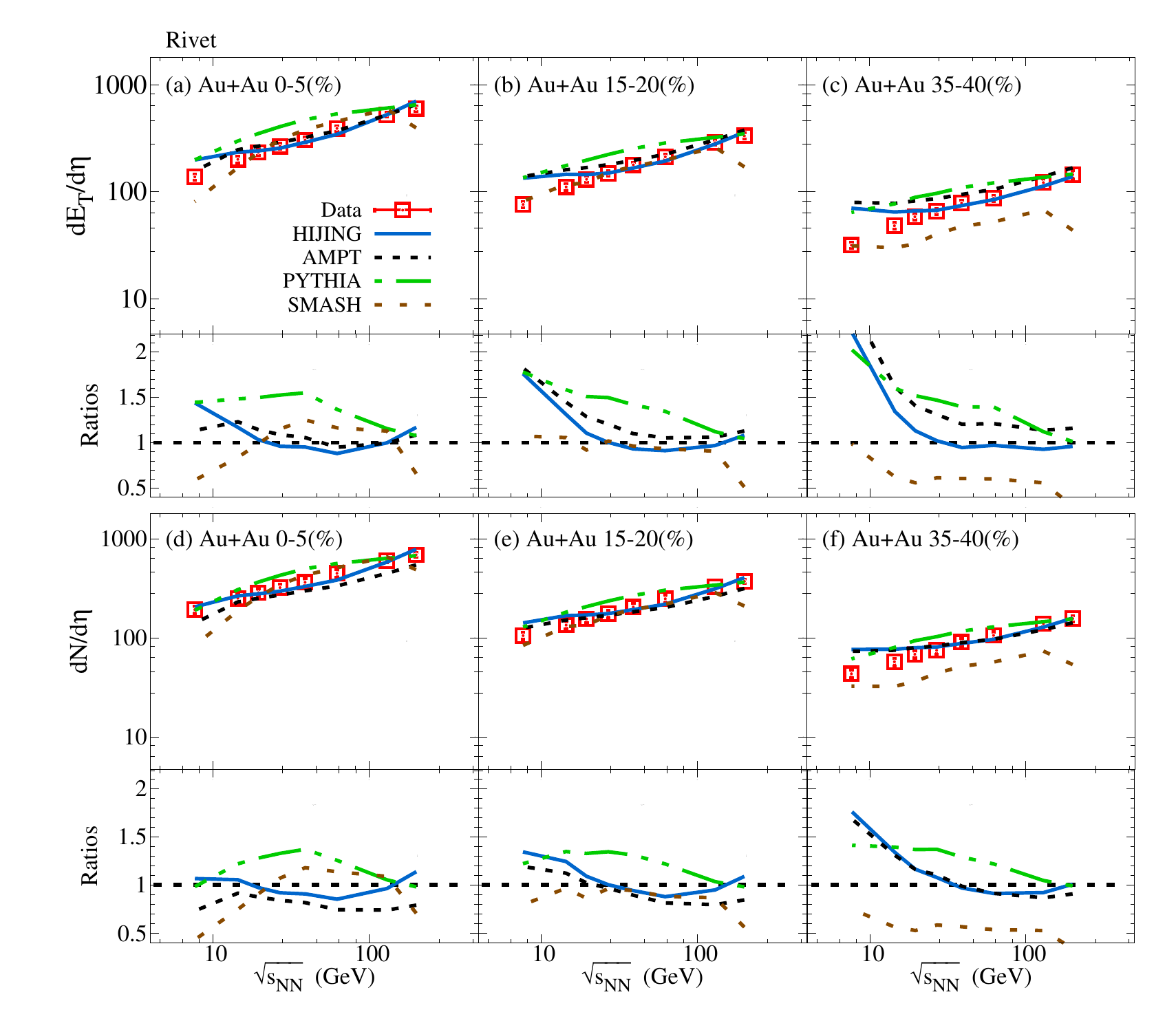}
\vskip -0.36cm
 \caption{
 Comparison of the beam energy dependence of $dE_{T}/d\eta$ and $dN/d\eta$ for Au+Au collisions and for centrally selections 0--5\%,  15--20\%, and  35--40\%.  The ratios are given relative to the experimental data. 
 \label{fig:02}
 }
 \vskip -0.46cm
 }
\end{figure*}

The size and transverse shape of the interaction region in nucleus-nucleus collisions cannot be measured directly. Experiments estimate centrality using a single event observable $N$, defining centrality in percentiles of the inelastic cross-section distribution ${d\sigma_{\text{inel}}}/{dN}$. Usually, $N$ is the charged particle multiplicity, although sometimes the transverse energy is used.  The centrality $c$ of a collision is expressed mathematically as:
\begin{equation}
	\label{eq:centrality}
	c = \frac{1}{\sigma_{\mathrm{inel}}} \int_N^\infty \frac{\mathrm{d}\sigma_{\mathrm{inel}}}{\mathrm{d}N'} \mathrm{d}N'.
\end{equation}

In Rivet, the observable $N$ can be defined in several ways~\cite{Bierlich:2020wms}.  The closest analog to the experimentally measured observable (such as the charged particle multiplicity) can be used directly, with percentiles translating to the corresponding cuts in $N$. This method aligns closely with Rivet's design philosophy but works primarily when the generator simulates a realistic distribution of collisions.  Alternately, a centrality percentile can be defined by the generator in the HEPMC output.  This can be useful when the generator does not simulate a full, realistic distribution of collisions.  Finally, the centrality percentile can be defined as a percentile of the impact parameter distribution.  This may be appropriate if the full distribution of collision centralities is simulated, but particles are either not simulated for the entire phase space, or the simulation does not produce reliable predictions for $N$ in the same region used by the experiment.

In many experiments, \(N\) represents an observable related to particle production in the forward region. To ensure flexibility and avoid restricting Monte Carlo generators, Rivet supports all three centrality definitions. Prior to using Rivet for data-model comparisons, a calibration run must be performed for each collision energy and system. This run generates the necessary histograms for \(\frac{d\sigma_{\text{inel}}}{dN}\) and \(\frac{d\sigma_{\text{inel}}}{db}\), which Rivet then uses to estimate centrality.

In PHENIX, the beam-beam counter (BBC) is used to determine centrality.  The BBC is a scintillator detector located at forward rapidities.  Following the method in Ref.~\cite{Bierlich:2020wms}, we implemented a centrality calibration analysis for the PHENIX experiment which records the multiplicities in the active region of the BBC ($3.1< | \eta| < 3.9$) and determines multiplicity thresholds to define centralities in the model as a percentile of the total collision cross section.  Each collision system and energy is calibrated separately for every model.
\section{RESULTS}\label{sec:3}
\begin{figure*}[t]
 \centering{
 \includegraphics[width=\textwidth,angle=0]{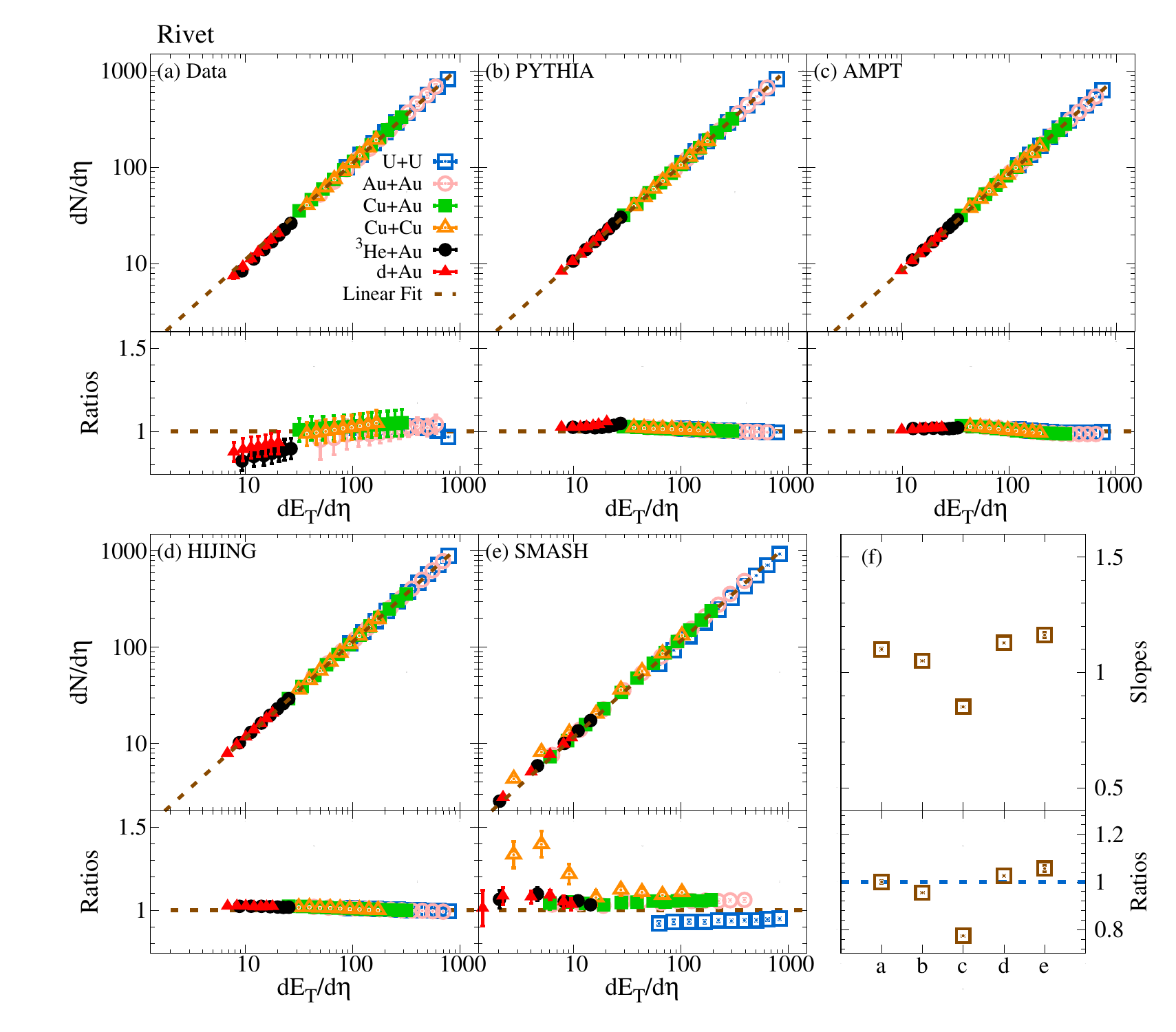}
\vskip -0.36cm
 \caption{
The system-size dependence of the measured correlations between \(dN/d\eta\) and \(dE_{T}/d\eta\) is compared between experimental data and various models, along with the ratios of the data points to the corresponding fits shown in panels (a)–(e). The slopes of the fits and their ratios relative to the data fit are presented in panel (f).}
 \label{fig:03}
 \vskip -0.46cm
 }
\end{figure*}

We compare charged particle multiplicities \(dN/d\eta\) and transverse energies \(dE_{T}/d\eta\) across various collision energies and system sizes, summarized in Table~\ref{tab:1}. Figure~\ref{fig:01} shows the multiplicity distributions calculated by the AMPT, HIJING, PYTHIA-8, and SMASH models for several collision energies and system sizes.  Most models have similar shapes at low multiplicities, but the distributions have different shapes at high multiplicities.  Since the centrality is defined as a percentile of the cross section, models with more events at high charged particle multiplicities will have higher multiplicities and higher transverse energies for the same centrality percentile. Our simulations reveal an absence of a consistent trend in model calculations as we move from larger to smaller collision energies and system sizes. This variability is expected to influence the different models' centrality definitions and, consequently, the centrality dependence of \(dN/d\eta\) and \(dE_{T}/d\eta\), highlighting the complexities of modeling these observables under different experimental conditions. 

Figure~\ref{fig:02} illustrates the beam energy dependence of \(dE_{T}/d\eta\) (panels (a)--(f)) and \(dN/d\eta\) (panels (g)--(l)) for three centrality selections: 0--5\%, 15--20\%, and 35--40\%. The experimental measurements demonstrate a monotonic decrease in both \(dE_{T}/d\eta\) and \(dN/d\eta\) with decreasing beam energy. 
This behavior is attributed to reduced particle and energy production at lower energies in the center-of-mass system. As the beam energy decreases, less energy becomes available for particle creation, leading to fewer produced particles and reduced transverse energy at midrapidity. This reduction stems from lower initial temperature and energy density in the collisions. Additionally, the increased nuclear-stopping power at lower beam energies results in more localized energy deposition near the collision region. This limits forward and backward longitudinal expansion, narrows the pseudorapidity distributions, and lowers the yields of both charged particles and transverse energy at midrapidity. The data model comparisons for different collision beam energies are given in detail in Appendix~\ref{Apx1}. 

Figure~\ref{fig:03} shows the correlation between  \(dE_{T}/d\eta\) and \(dN/d\eta\) for various collision systems at $\sqrt{s_{\mathrm{NN}}}\sim200$~GeV, ranging from small systems like d+Au and $^{3}$He+Au to larger systems such as U+U and Au+Au from data and models. The data points align well with a linear fit, indicating a direct proportionality between the two observables. This linear behavior suggests that both  \(dE_{T}/d\eta\) and \(dN/d\eta\) scale similarly with the size of the colliding system, both in data and models. While the ratios remain close to unity, deviations of about 10\% are observed, particularly for small systems. The overall consistency of the ratios with the linear fit supports the interpretation of a common scaling mechanism, reinforcing the idea that both observables are influenced by similar physics processes from small to large systems in data and models. 
Fig.~\ref{fig:03}(f) compares the slopes of fits to a straight line between data and the models.  The slopes for data, PYTHIA, HIJING, and SMASH agree within 5\% while AMPT is lower than the data by about 25\%.
In addition, the comparisons of the fit slopes shown in Fig.~\ref{fig:03}(f?) indicate that, except for the AMPT model, which deviates from the data fit by approximately 25\%, the other models agree with the data behavior within about 5\%.
The data-model comparisons for different collision systems are shown in detail in Appendix~\ref{Apx1}. 

HIJING and AMPT use PYTHIA-6 with a Glauber model for the initial state.  AMPT adds elastic parton scattering, recombination, and hadronic rescattering.  PYTHIA Angantyr uses PYTHIA-8 for nucleon-nucleon scattering combined with a Glauber model, while additing some additional multiparton interactions to model nucleus-nucleus interactions.  
While PYTHIA-8 and HIJING focus on initial hard and soft scattering processes combined with string fragmentation, AMPT incorporates a multi-phase transport framework that includes both partonic and hadronic stages. 
SMASH produces strings using its own excitation routine, gradually switching to PYTHIA's HardQCD processes as energy increases~\cite{Mohs:2019iee}
These differences highlight each model’s distinct physics emphases and their respective applications in simulating particle production in nuclear collisions.   The ability of these models to describe the data with comparable fidelity may indicate either that these models have been tuned extensively or that more detailed tests are required to distinguish models.
Although the models exhibit trends similar to the experimental measurements shown in Figs.~\ref{fig:02} and~\ref{fig:03} across various energies and system sizes, quantitatively reproducing the data at low energies and in peripheral collisions remains challenging. Notably, SMASH displays the most significant deviations from experimental results in peripheral collisions at lower beam energies, emphasizing unique challenges inherent to its modeling approach.
Model-to-data ratios (see Appendix \ref{Apx1}) exhibit the largest discrepancies in these regions, highlighting limitations in the current theoretical descriptions. These results indicate that all models require additional constraints or modifications to achieve better agreement with experimental data. Moreover, future studies employing the Rivet framework for more sophisticated observables, such as particle correlations, will be essential for further distinguishing between theoretical approaches.

\section{CONCLUSION}\label{sec:4}
We use the Rivet framework to present comprehensive data model comparisons of \(dE_{T}/d\eta\) and \(dN/d\eta\) in midrapidity for several collision systems and beam energies.  Our data model comparisons indicate that while the PYTHIA-8, AMPT, HIJING, and SMASH models successfully describe some features of the experimental data, they struggle with precise predictions in low-energy and peripheral collisions. Discrepancies in model-to-data ratios highlight the need for refined theoretical approaches, particularly in modeling baryon stopping and initial-state dynamics.  We observed linear scaling between \(dE_{T}/d\eta\) and \(dN/d\eta\)  across all systems, both in data and models. This observation suggests that energy deposition and particle production share a common underlying mechanism tied to the number of participant nucleons. Because the data and various models agree to a similar extent, future studies using the Rivet framework to examine more refined observables will be essential for distinguishing between theoretical approaches.

\section*{Acknowledgments}
This work was supported in part by funding from the Division of Nuclear Physics of the U.S. Department of Energy under Grant No. DE-FG02-96ER40982 and from the National Science Foundation under Grant No. OAC-1550300.  We are grateful to Steffen Bass for productive discussions, and to Carl Rosenkvist and Hannah Elfner for SMASH calculations and useful discussions.

\appendix
\begin{widetext}

\section{Centrality dependence}\label{Apx1}

In this appendix, we provide the complete set of data as a function of collision centrality, corresponding to the observables presented in the main text. 
\begin{figure}[!h]
 \centering{
 \includegraphics[width=\textwidth,angle=0]{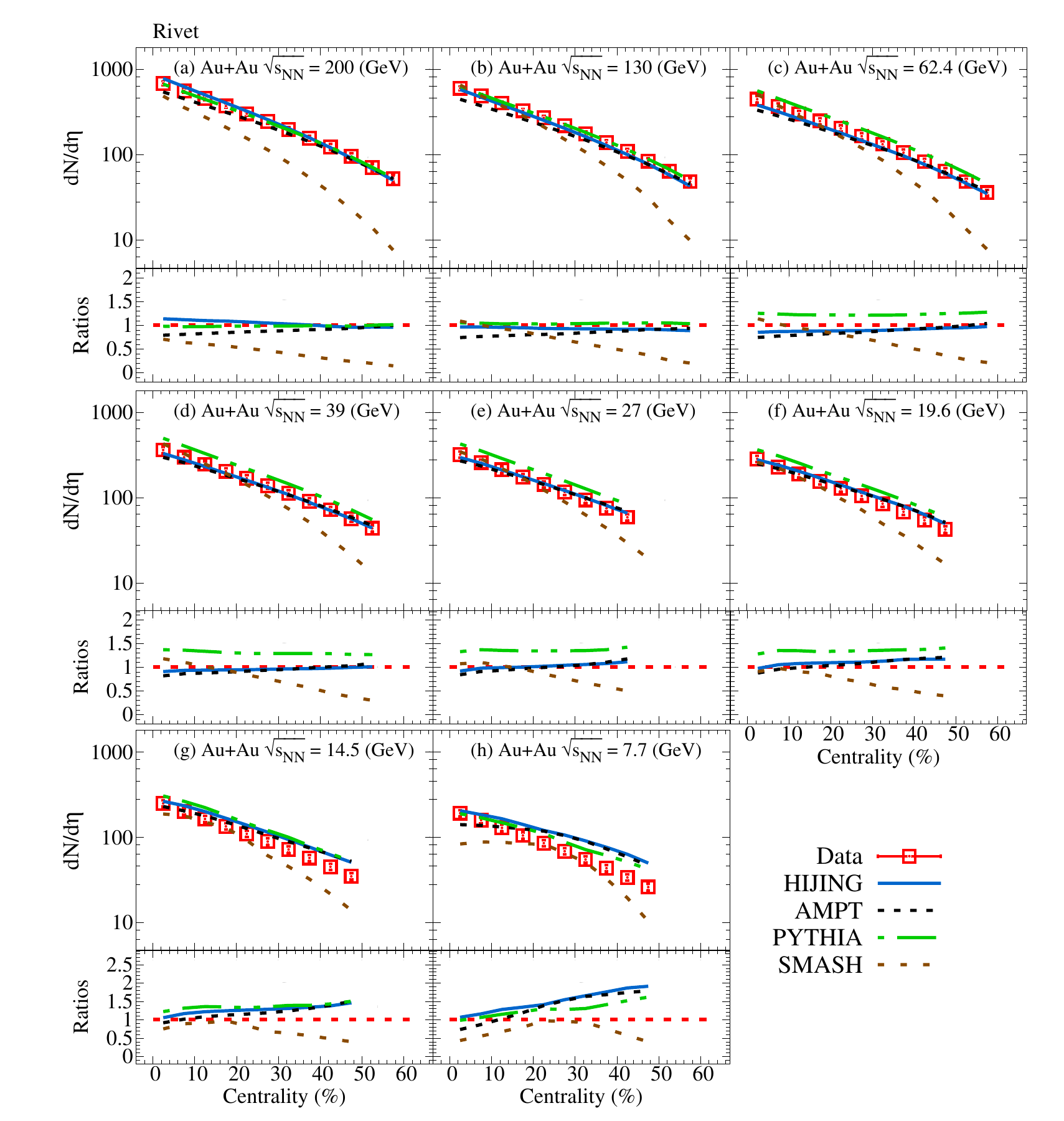}
\vskip -0.26cm
 \caption{
 Comparison of the $dN/d\eta$ centrally dependence for Au+Au at $\sqrt(s_{NN})$ = 7.7--200 GeV. The lines represent the MC model's calculations.
 \label{fig:05}
 }
 \vskip -0.46cm
 }
\end{figure}

\begin{figure}[!h]
 \centering{
 \includegraphics[width=\textwidth,angle=0]{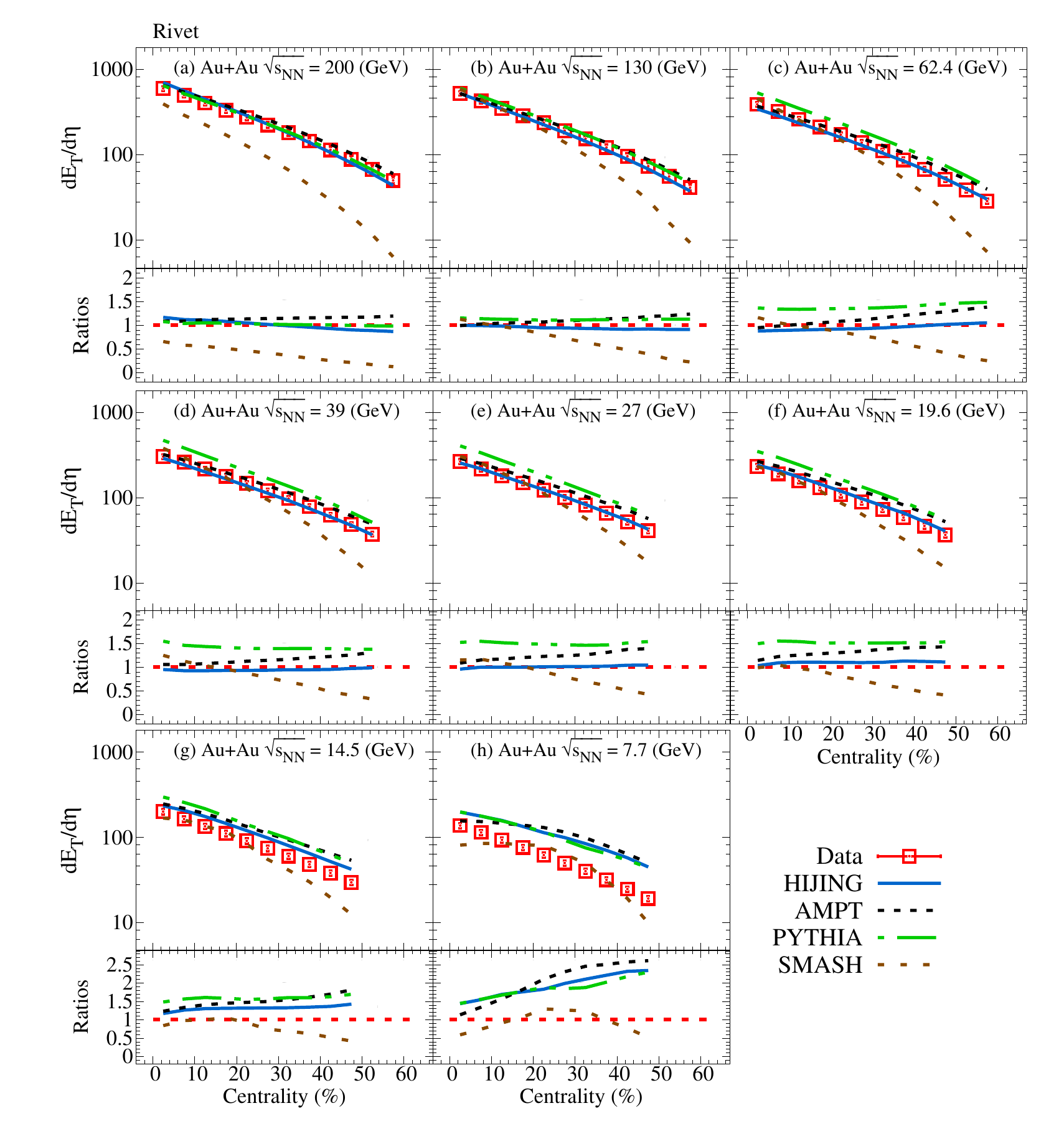}
\vskip -0.26cm
 \caption{
 Comparison of the $dE_{T}/d\eta$ centrally dependence for Au+Au at $\sqrt(s_{NN})$ = 7.7--200 GeV. The lines represent the MC model's calculations.
 \label{fig:06}
 }
 \vskip -0.46cm
 }
\end{figure}

\begin{figure}[!h]
 \centering{
 \includegraphics[width=\textwidth,angle=0]{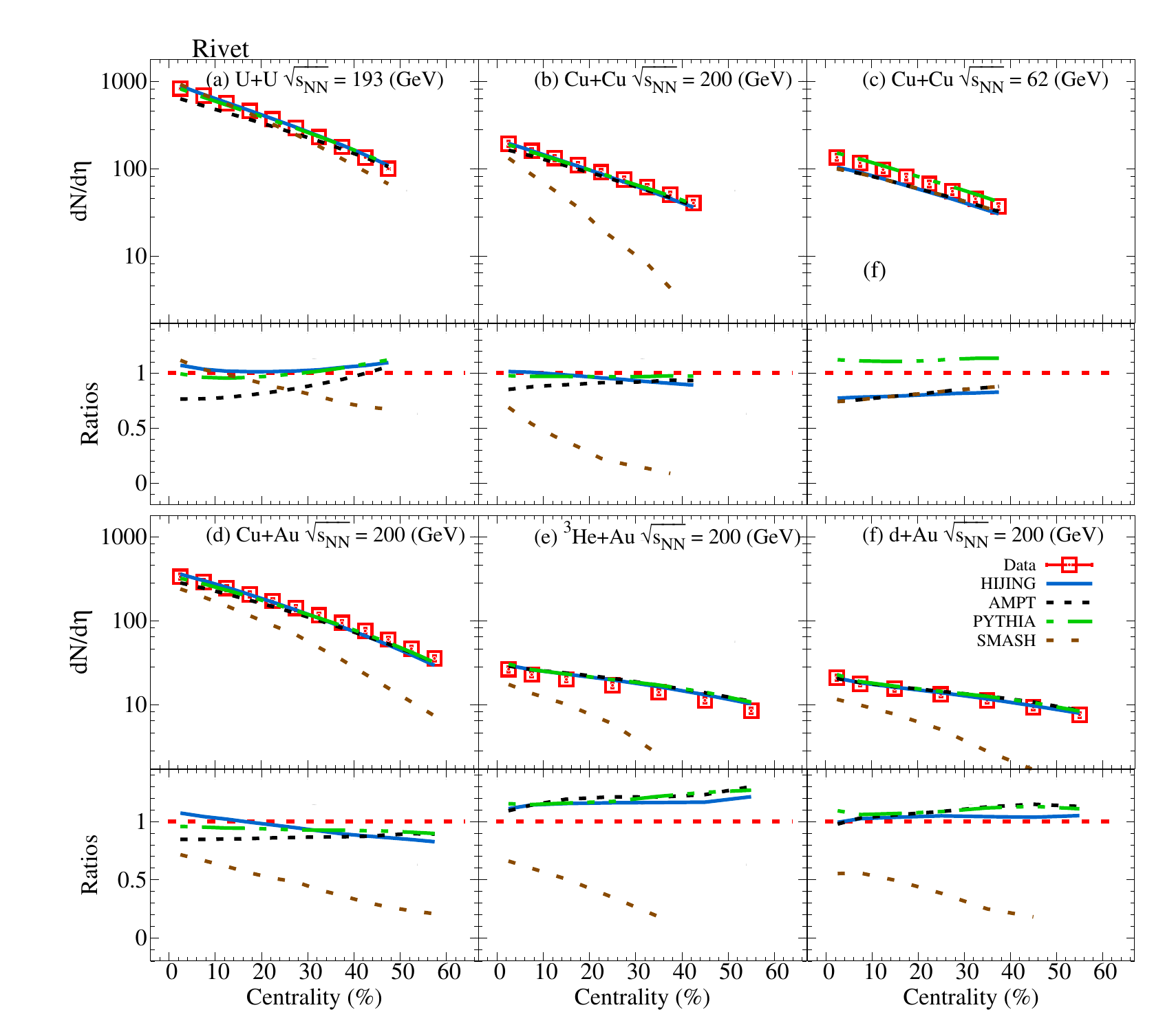}
\vskip -0.26cm
 \caption{
 Comparison of the $dN/d\eta$ central dependence for U+U at 193, Cu+Au at 62.4 and 200 GeV, Cu+Au, $^3$He+Au and d+Au at 200 GeV. The lines represent the MC model's calculations.
 \label{fig:07}
 }
 \vskip -0.46cm
 }
\end{figure}

\begin{figure}[!h]
 \centering{
 \includegraphics[width=\textwidth,angle=0]{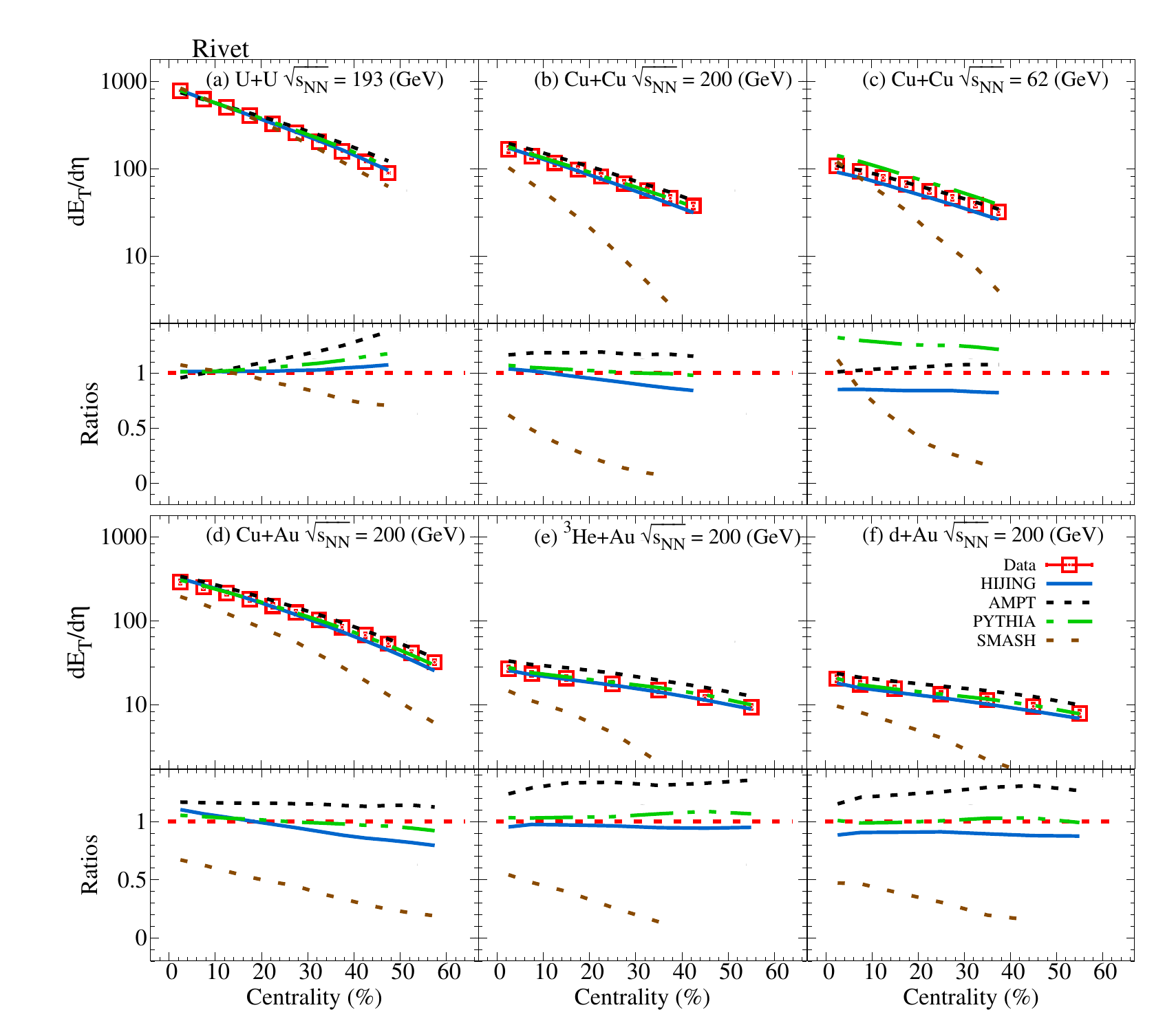}
\vskip -0.26cm
 \caption{
 Comparison of the $dE_{T}/d\eta$ central dependence for U+U at 193, Cu+Au at 62.4 and 200 GeV, Cu+Au, $^3$He+Au and d+Au at 200 GeV. The lines represent the MC model's calculations.
 \label{fig:08}
 }
 \vskip -0.46cm
 }
\end{figure}

\end{widetext}

\clearpage
\bibliography{ref} 

\begin{thebibliography}{95}%
\makeatletter
\providecommand \@ifxundefined [1]{%
 \@ifx{#1\undefined}
}%
\providecommand \@ifnum [1]{%
 \ifnum #1\expandafter \@firstoftwo
 \else \expandafter \@secondoftwo
 \fi
}%
\providecommand \@ifx [1]{%
 \ifx #1\expandafter \@firstoftwo
 \else \expandafter \@secondoftwo
 \fi
}%
\providecommand \natexlab [1]{#1}%
\providecommand \enquote  [1]{``#1''}%
\providecommand \bibnamefont  [1]{#1}%
\providecommand \bibfnamefont [1]{#1}%
\providecommand \citenamefont [1]{#1}%
\providecommand \href@noop [0]{\@secondoftwo}%
\providecommand \href [0]{\begingroup \@sanitize@url \@href}%
\providecommand \@href[1]{\@@startlink{#1}\@@href}%
\providecommand \@@href[1]{\endgroup#1\@@endlink}%
\providecommand \@sanitize@url [0]{\catcode `\\12\catcode `\$12\catcode
  `\&12\catcode `\#12\catcode `\^12\catcode `\_12\catcode `\%12\relax}%
\providecommand \@@startlink[1]{}%
\providecommand \@@endlink[0]{}%
\providecommand \url  [0]{\begingroup\@sanitize@url \@url }%
\providecommand \@url [1]{\endgroup\@href {#1}{\urlprefix }}%
\providecommand \urlprefix  [0]{URL }%
\providecommand \Eprint [0]{\href }%
\providecommand \doibase [0]{http://dx.doi.org/}%
\providecommand \selectlanguage [0]{\@gobble}%
\providecommand \bibinfo  [0]{\@secondoftwo}%
\providecommand \bibfield  [0]{\@secondoftwo}%
\providecommand \translation [1]{[#1]}%
\providecommand \BibitemOpen [0]{}%
\providecommand \bibitemStop [0]{}%
\providecommand \bibitemNoStop [0]{.\EOS\space}%
\providecommand \EOS [0]{\spacefactor3000\relax}%
\providecommand \BibitemShut  [1]{\csname bibitem#1\endcsname}%
\let\auto@bib@innerbib\@empty
\bibitem [{\citenamefont {Heinz}\ and\ \citenamefont
  {Jacob}(2000)}]{Heinz:2000bk}%
  \BibitemOpen
  \bibfield  {author} {\bibinfo {author} {\bibfnamefont {U.~W.}\ \bibnamefont
  {Heinz}}\ and\ \bibinfo {author} {\bibfnamefont {M.}~\bibnamefont {Jacob}},\
  }\href@noop {} {\  (\bibinfo {year} {2000})},\ \Eprint
  {http://arxiv.org/abs/nucl-th/0002042} {arXiv:nucl-th/0002042} \BibitemShut
  {NoStop}%
\bibitem [{\citenamefont {Busza}\ \emph {et~al.}(2018)\citenamefont {Busza},
  \citenamefont {Rajagopal},\ and\ \citenamefont {van~der
  Schee}}]{Busza:2018rrf}%
  \BibitemOpen
  \bibfield  {author} {\bibinfo {author} {\bibfnamefont {W.}~\bibnamefont
  {Busza}}, \bibinfo {author} {\bibfnamefont {K.}~\bibnamefont {Rajagopal}}, \
  and\ \bibinfo {author} {\bibfnamefont {W.}~\bibnamefont {van~der Schee}},\
  }\href {\doibase 10.1146/annurev-nucl-101917-020852} {\bibfield  {journal}
  {\bibinfo  {journal} {Ann. Rev. Nucl. Part. Sci.}\ }\textbf {\bibinfo
  {volume} {68}},\ \bibinfo {pages} {339} (\bibinfo {year} {2018})},\ \Eprint
  {http://arxiv.org/abs/1802.04801} {arXiv:1802.04801 [hep-ph]} \BibitemShut
  {NoStop}%
\bibitem [{\citenamefont {Heinz}\ and\ \citenamefont
  {Snellings}(2013)}]{Heinz:2013th}%
  \BibitemOpen
  \bibfield  {author} {\bibinfo {author} {\bibfnamefont {U.}~\bibnamefont
  {Heinz}}\ and\ \bibinfo {author} {\bibfnamefont {R.}~\bibnamefont
  {Snellings}},\ }\href {\doibase 10.1146/annurev-nucl-102212-170540}
  {\bibfield  {journal} {\bibinfo  {journal} {Ann. Rev. Nucl. Part. Sci.}\
  }\textbf {\bibinfo {volume} {63}},\ \bibinfo {pages} {123} (\bibinfo {year}
  {2013})}\BibitemShut {NoStop}%
\bibitem [{\citenamefont {Shuryak}(1978)}]{Shuryak:1978ij}%
  \BibitemOpen
  \bibfield  {author} {\bibinfo {author} {\bibfnamefont {E.~V.}\ \bibnamefont
  {Shuryak}},\ }\href {\doibase 10.1016/0370-2693(78)90370-2} {\bibfield
  {journal} {\bibinfo  {journal} {Sov. J. Nucl. Phys.}\ }\textbf {\bibinfo
  {volume} {28}},\ \bibinfo {pages} {408} (\bibinfo {year} {1978})}\BibitemShut
  {NoStop}%
\bibitem [{\citenamefont {Shuryak}(1980)}]{Shuryak:1980tp}%
  \BibitemOpen
  \bibfield  {author} {\bibinfo {author} {\bibfnamefont {E.~V.}\ \bibnamefont
  {Shuryak}},\ }\href {\doibase 10.1016/0370-1573(80)90105-2} {\bibfield
  {journal} {\bibinfo  {journal} {Phys. Rept.}\ }\textbf {\bibinfo {volume}
  {61}},\ \bibinfo {pages} {71} (\bibinfo {year} {1980})}\BibitemShut {NoStop}%
\bibitem [{\citenamefont {Muller}\ \emph {et~al.}(2012)\citenamefont {Muller},
  \citenamefont {Schukraft},\ and\ \citenamefont {Wyslouch}}]{Muller:2012zq}%
  \BibitemOpen
  \bibfield  {author} {\bibinfo {author} {\bibfnamefont {B.}~\bibnamefont
  {Muller}}, \bibinfo {author} {\bibfnamefont {J.}~\bibnamefont {Schukraft}}, \
  and\ \bibinfo {author} {\bibfnamefont {B.}~\bibnamefont {Wyslouch}},\ }\href
  {\doibase 10.1146/annurev-nucl-102711-094910} {\bibfield  {journal} {\bibinfo
   {journal} {Ann. Rev. Nucl. Part. Sci.}\ }\textbf {\bibinfo {volume} {62}},\
  \bibinfo {pages} {361} (\bibinfo {year} {2012})},\ \Eprint
  {http://arxiv.org/abs/1202.3233} {arXiv:1202.3233 [hep-ex]} \BibitemShut
  {NoStop}%
\bibitem [{\citenamefont {Adams}\ \emph {et~al.}(2005)\citenamefont {Adams}
  \emph {et~al.}}]{STAR:2005gfr}%
  \BibitemOpen
  \bibfield  {author} {\bibinfo {author} {\bibfnamefont {J.}~\bibnamefont
  {Adams}} \emph {et~al.} (\bibinfo {collaboration} {STAR}),\ }\href {\doibase
  10.1016/j.nuclphysa.2005.03.085} {\bibfield  {journal} {\bibinfo  {journal}
  {Nucl. Phys. A}\ }\textbf {\bibinfo {volume} {757}},\ \bibinfo {pages} {102}
  (\bibinfo {year} {2005})},\ \Eprint {http://arxiv.org/abs/nucl-ex/0501009}
  {arXiv:nucl-ex/0501009} \BibitemShut {NoStop}%
\bibitem [{\citenamefont {Adcox}\ \emph {et~al.}(2005)\citenamefont {Adcox}
  \emph {et~al.}}]{PHENIX:2004vcz}%
  \BibitemOpen
  \bibfield  {author} {\bibinfo {author} {\bibfnamefont {K.}~\bibnamefont
  {Adcox}} \emph {et~al.} (\bibinfo {collaboration} {PHENIX}),\ }\href
  {\doibase 10.1016/j.nuclphysa.2005.03.086} {\bibfield  {journal} {\bibinfo
  {journal} {Nucl. Phys. A}\ }\textbf {\bibinfo {volume} {757}},\ \bibinfo
  {pages} {184} (\bibinfo {year} {2005})},\ \Eprint
  {http://arxiv.org/abs/nucl-ex/0410003} {arXiv:nucl-ex/0410003} \BibitemShut
  {NoStop}%
\bibitem [{\citenamefont {Arsene}\ \emph {et~al.}(2005)\citenamefont {Arsene}
  \emph {et~al.}}]{BRAHMS:2004adc}%
  \BibitemOpen
  \bibfield  {author} {\bibinfo {author} {\bibfnamefont {I.}~\bibnamefont
  {Arsene}} \emph {et~al.} (\bibinfo {collaboration} {BRAHMS}),\ }\href
  {\doibase 10.1016/j.nuclphysa.2005.02.130} {\bibfield  {journal} {\bibinfo
  {journal} {Nucl. Phys. A}\ }\textbf {\bibinfo {volume} {757}},\ \bibinfo
  {pages} {1} (\bibinfo {year} {2005})},\ \Eprint
  {http://arxiv.org/abs/nucl-ex/0410020} {arXiv:nucl-ex/0410020} \BibitemShut
  {NoStop}%
\bibitem [{\citenamefont {Back}\ \emph {et~al.}(2005)\citenamefont {Back} \emph
  {et~al.}}]{PHOBOS:2004zne}%
  \BibitemOpen
  \bibfield  {author} {\bibinfo {author} {\bibfnamefont {B.~B.}\ \bibnamefont
  {Back}} \emph {et~al.} (\bibinfo {collaboration} {PHOBOS}),\ }\href {\doibase
  10.1016/j.nuclphysa.2005.03.084} {\bibfield  {journal} {\bibinfo  {journal}
  {Nucl. Phys. A}\ }\textbf {\bibinfo {volume} {757}},\ \bibinfo {pages} {28}
  (\bibinfo {year} {2005})},\ \Eprint {http://arxiv.org/abs/nucl-ex/0410022}
  {arXiv:nucl-ex/0410022} \BibitemShut {NoStop}%
\bibitem [{\citenamefont {Adare}\ \emph {et~al.}(2016)\citenamefont {Adare}
  \emph {et~al.}}]{PHENIX:2015tbb}%
  \BibitemOpen
  \bibfield  {author} {\bibinfo {author} {\bibfnamefont {A.}~\bibnamefont
  {Adare}} \emph {et~al.} (\bibinfo {collaboration} {PHENIX}),\ }\href
  {\doibase 10.1103/PhysRevC.93.024901} {\bibfield  {journal} {\bibinfo
  {journal} {Phys. Rev. C}\ }\textbf {\bibinfo {volume} {93}},\ \bibinfo
  {pages} {024901} (\bibinfo {year} {2016})},\ \Eprint
  {http://arxiv.org/abs/1509.06727} {arXiv:1509.06727 [nucl-ex]} \BibitemShut
  {NoStop}%
\bibitem [{\citenamefont {Adler}\ \emph {et~al.}(2005)\citenamefont {Adler}
  \emph {et~al.}}]{PHENIX:2004vdg}%
  \BibitemOpen
  \bibfield  {author} {\bibinfo {author} {\bibfnamefont {S.~S.}\ \bibnamefont
  {Adler}} \emph {et~al.} (\bibinfo {collaboration} {PHENIX}),\ }\href
  {\doibase 10.1103/PhysRevC.71.034908} {\bibfield  {journal} {\bibinfo
  {journal} {Phys. Rev. C}\ }\textbf {\bibinfo {volume} {71}},\ \bibinfo
  {pages} {034908} (\bibinfo {year} {2005})},\ \bibinfo {note} {[Erratum:
  Phys.Rev.C 71, 049901 (2005)]},\ \Eprint
  {http://arxiv.org/abs/nucl-ex/0409015} {arXiv:nucl-ex/0409015} \BibitemShut
  {NoStop}%
\bibitem [{\citenamefont {Cleymans}\ \emph {et~al.}(2008)\citenamefont
  {Cleymans}, \citenamefont {Sahoo}, \citenamefont {Mahapatra}, \citenamefont
  {Srivastava},\ and\ \citenamefont {Wheaton}}]{Cleymans:2007uk}%
  \BibitemOpen
  \bibfield  {author} {\bibinfo {author} {\bibfnamefont {J.}~\bibnamefont
  {Cleymans}}, \bibinfo {author} {\bibfnamefont {R.}~\bibnamefont {Sahoo}},
  \bibinfo {author} {\bibfnamefont {D.~P.}\ \bibnamefont {Mahapatra}}, \bibinfo
  {author} {\bibfnamefont {D.~K.}\ \bibnamefont {Srivastava}}, \ and\ \bibinfo
  {author} {\bibfnamefont {S.}~\bibnamefont {Wheaton}},\ }\href {\doibase
  10.1016/j.physletb.2007.12.029} {\bibfield  {journal} {\bibinfo  {journal}
  {Phys. Lett. B}\ }\textbf {\bibinfo {volume} {660}},\ \bibinfo {pages} {172}
  (\bibinfo {year} {2008})},\ \Eprint {http://arxiv.org/abs/0708.0914}
  {arXiv:0708.0914 [hep-ph]} \BibitemShut {NoStop}%
\bibitem [{\citenamefont {Bozek}(2005)}]{Bozek:2005eu}%
  \BibitemOpen
  \bibfield  {author} {\bibinfo {author} {\bibfnamefont {P.}~\bibnamefont
  {Bozek}},\ }\href@noop {} {\bibfield  {journal} {\bibinfo  {journal} {Acta
  Phys. Polon. B}\ }\textbf {\bibinfo {volume} {36}},\ \bibinfo {pages} {3071}
  (\bibinfo {year} {2005})},\ \Eprint {http://arxiv.org/abs/nucl-th/0506037}
  {arXiv:nucl-th/0506037} \BibitemShut {NoStop}%
\bibitem [{\citenamefont {Kharzeev}\ \emph {et~al.}(2005)\citenamefont
  {Kharzeev}, \citenamefont {Levin},\ and\ \citenamefont
  {Nardi}}]{Kharzeev:2001yq}%
  \BibitemOpen
  \bibfield  {author} {\bibinfo {author} {\bibfnamefont {D.}~\bibnamefont
  {Kharzeev}}, \bibinfo {author} {\bibfnamefont {E.}~\bibnamefont {Levin}}, \
  and\ \bibinfo {author} {\bibfnamefont {M.}~\bibnamefont {Nardi}},\ }\href
  {\doibase 10.1103/PhysRevC.71.054903} {\bibfield  {journal} {\bibinfo
  {journal} {Phys. Rev. C}\ }\textbf {\bibinfo {volume} {71}},\ \bibinfo
  {pages} {054903} (\bibinfo {year} {2005})},\ \Eprint
  {http://arxiv.org/abs/hep-ph/0111315} {arXiv:hep-ph/0111315} \BibitemShut
  {NoStop}%
\bibitem [{\citenamefont {Mendenhall}\ and\ \citenamefont
  {Lin}(2021)}]{Mendenhall:2020fil}%
  \BibitemOpen
  \bibfield  {author} {\bibinfo {author} {\bibfnamefont {T.}~\bibnamefont
  {Mendenhall}}\ and\ \bibinfo {author} {\bibfnamefont {Z.-W.}\ \bibnamefont
  {Lin}},\ }\href {\doibase 10.1103/PhysRevC.103.024907} {\bibfield  {journal}
  {\bibinfo  {journal} {Phys. Rev. C}\ }\textbf {\bibinfo {volume} {103}},\
  \bibinfo {pages} {024907} (\bibinfo {year} {2021})},\ \Eprint
  {http://arxiv.org/abs/2012.13825} {arXiv:2012.13825 [nucl-th]} \BibitemShut
  {NoStop}%
\bibitem [{\citenamefont {Auvinen}\ \emph {et~al.}(2020)\citenamefont
  {Auvinen}, \citenamefont {Eskola}, \citenamefont {Huovinen}, \citenamefont
  {Niemi}, \citenamefont {Paatelainen},\ and\ \citenamefont
  {Petreczky}}]{Auvinen:2020mpc}%
  \BibitemOpen
  \bibfield  {author} {\bibinfo {author} {\bibfnamefont {J.}~\bibnamefont
  {Auvinen}}, \bibinfo {author} {\bibfnamefont {K.~J.}\ \bibnamefont {Eskola}},
  \bibinfo {author} {\bibfnamefont {P.}~\bibnamefont {Huovinen}}, \bibinfo
  {author} {\bibfnamefont {H.}~\bibnamefont {Niemi}}, \bibinfo {author}
  {\bibfnamefont {R.}~\bibnamefont {Paatelainen}}, \ and\ \bibinfo {author}
  {\bibfnamefont {P.}~\bibnamefont {Petreczky}},\ }\href {\doibase
  10.1103/PhysRevC.102.044911} {\bibfield  {journal} {\bibinfo  {journal}
  {Phys. Rev. C}\ }\textbf {\bibinfo {volume} {102}},\ \bibinfo {pages}
  {044911} (\bibinfo {year} {2020})},\ \Eprint
  {http://arxiv.org/abs/2006.12499} {arXiv:2006.12499 [nucl-th]} \BibitemShut
  {NoStop}%
\bibitem [{\citenamefont {Karpenko}\ \emph {et~al.}(2015)\citenamefont
  {Karpenko}, \citenamefont {Huovinen}, \citenamefont {Petersen},\ and\
  \citenamefont {Bleicher}}]{Karpenko:2015xea}%
  \BibitemOpen
  \bibfield  {author} {\bibinfo {author} {\bibfnamefont {I.~A.}\ \bibnamefont
  {Karpenko}}, \bibinfo {author} {\bibfnamefont {P.}~\bibnamefont {Huovinen}},
  \bibinfo {author} {\bibfnamefont {H.}~\bibnamefont {Petersen}}, \ and\
  \bibinfo {author} {\bibfnamefont {M.}~\bibnamefont {Bleicher}},\ }\href
  {\doibase 10.1103/PhysRevC.91.064901} {\bibfield  {journal} {\bibinfo
  {journal} {Phys. Rev. C}\ }\textbf {\bibinfo {volume} {91}},\ \bibinfo
  {pages} {064901} (\bibinfo {year} {2015})}\BibitemShut {NoStop}%
\bibitem [{\citenamefont {Jahan}\ \emph {et~al.}(2024)\citenamefont {Jahan},
  \citenamefont {Roch},\ and\ \citenamefont {Shen}}]{Jahan:2024wpj}%
  \BibitemOpen
  \bibfield  {author} {\bibinfo {author} {\bibfnamefont {S.~A.}\ \bibnamefont
  {Jahan}}, \bibinfo {author} {\bibfnamefont {H.}~\bibnamefont {Roch}}, \ and\
  \bibinfo {author} {\bibfnamefont {C.}~\bibnamefont {Shen}},\ }\href {\doibase
  10.1103/PhysRevC.110.054905} {\bibfield  {journal} {\bibinfo  {journal}
  {Phys. Rev. C}\ }\textbf {\bibinfo {volume} {110}},\ \bibinfo {pages}
  {054905} (\bibinfo {year} {2024})},\ \Eprint
  {http://arxiv.org/abs/2408.00537} {arXiv:2408.00537 [nucl-th]} \BibitemShut
  {NoStop}%
\bibitem [{\citenamefont {Adare}\ \emph {et~al.}(2013)\citenamefont {Adare}
  \emph {et~al.}}]{PHENIX:2013ktj}%
  \BibitemOpen
  \bibfield  {author} {\bibinfo {author} {\bibfnamefont {A.}~\bibnamefont
  {Adare}} \emph {et~al.} (\bibinfo {collaboration} {PHENIX}),\ }\href
  {\doibase 10.1103/PhysRevLett.111.212301} {\bibfield  {journal} {\bibinfo
  {journal} {Phys. Rev. Lett.}\ }\textbf {\bibinfo {volume} {111}},\ \bibinfo
  {pages} {212301} (\bibinfo {year} {2013})},\ \Eprint
  {http://arxiv.org/abs/1303.1794} {arXiv:1303.1794 [nucl-ex]} \BibitemShut
  {NoStop}%
\bibitem [{\citenamefont {Abdulameer}\ \emph {et~al.}(2023)\citenamefont
  {Abdulameer} \emph {et~al.}}]{PHENIX:2022nht}%
  \BibitemOpen
  \bibfield  {author} {\bibinfo {author} {\bibfnamefont {N.~J.}\ \bibnamefont
  {Abdulameer}} \emph {et~al.} (\bibinfo {collaboration} {PHENIX}),\ }\href
  {\doibase 10.1103/PhysRevC.107.024907} {\bibfield  {journal} {\bibinfo
  {journal} {Phys. Rev. C}\ }\textbf {\bibinfo {volume} {107}},\ \bibinfo
  {pages} {024907} (\bibinfo {year} {2023})},\ \Eprint
  {http://arxiv.org/abs/2203.09894} {arXiv:2203.09894 [nucl-ex]} \BibitemShut
  {NoStop}%
\bibitem [{\citenamefont {Abdulhamid}\ \emph {et~al.}(2023)\citenamefont
  {Abdulhamid} \emph {et~al.}}]{STAR:2022pfn}%
  \BibitemOpen
  \bibfield  {author} {\bibinfo {author} {\bibfnamefont {M.~I.}\ \bibnamefont
  {Abdulhamid}} \emph {et~al.} (\bibinfo {collaboration} {STAR}),\ }\href
  {\doibase 10.1103/PhysRevLett.130.242301} {\bibfield  {journal} {\bibinfo
  {journal} {Phys. Rev. Lett.}\ }\textbf {\bibinfo {volume} {130}},\ \bibinfo
  {pages} {242301} (\bibinfo {year} {2023})},\ \Eprint
  {http://arxiv.org/abs/2210.11352} {arXiv:2210.11352 [nucl-ex]} \BibitemShut
  {NoStop}%
\bibitem [{\citenamefont {Abdulhamid}\ \emph {et~al.}(2024)\citenamefont
  {Abdulhamid} \emph {et~al.}}]{STAR:2023wmd}%
  \BibitemOpen
  \bibfield  {author} {\bibinfo {author} {\bibfnamefont {M.~I.}\ \bibnamefont
  {Abdulhamid}} \emph {et~al.} (\bibinfo {collaboration} {STAR}),\ }\href
  {\doibase 10.1103/PhysRevC.110.064902} {\bibfield  {journal} {\bibinfo
  {journal} {Phys. Rev. C}\ }\textbf {\bibinfo {volume} {110}},\ \bibinfo
  {pages} {064902} (\bibinfo {year} {2024})},\ \Eprint
  {http://arxiv.org/abs/2312.07464} {arXiv:2312.07464 [nucl-ex]} \BibitemShut
  {NoStop}%
\bibitem [{\citenamefont {Aad}\ \emph {et~al.}(2016)\citenamefont {Aad} \emph
  {et~al.}}]{ATLAS:2015hzw}%
  \BibitemOpen
  \bibfield  {author} {\bibinfo {author} {\bibfnamefont {G.}~\bibnamefont
  {Aad}} \emph {et~al.} (\bibinfo {collaboration} {ATLAS}),\ }\href {\doibase
  10.1103/PhysRevLett.116.172301} {\bibfield  {journal} {\bibinfo  {journal}
  {Phys. Rev. Lett.}\ }\textbf {\bibinfo {volume} {116}},\ \bibinfo {pages}
  {172301} (\bibinfo {year} {2016})},\ \Eprint
  {http://arxiv.org/abs/1509.04776} {arXiv:1509.04776 [hep-ex]} \BibitemShut
  {NoStop}%
\bibitem [{\citenamefont {Chatrchyan}\ \emph {et~al.}(2013)\citenamefont
  {Chatrchyan} \emph {et~al.}}]{CMS:2012qk}%
  \BibitemOpen
  \bibfield  {author} {\bibinfo {author} {\bibfnamefont {S.}~\bibnamefont
  {Chatrchyan}} \emph {et~al.} (\bibinfo {collaboration} {CMS}),\ }\href
  {\doibase 10.1016/j.physletb.2012.11.025} {\bibfield  {journal} {\bibinfo
  {journal} {Phys. Lett.}\ }\textbf {\bibinfo {volume} {B718}},\ \bibinfo
  {pages} {795} (\bibinfo {year} {2013})},\ \Eprint
  {http://arxiv.org/abs/1210.5482} {arXiv:1210.5482 [nucl-ex]} \BibitemShut
  {NoStop}%
\bibitem [{\citenamefont {Abelev}\ \emph {et~al.}(2013)\citenamefont {Abelev}
  \emph {et~al.}}]{ALICE:2012eyl}%
  \BibitemOpen
  \bibfield  {author} {\bibinfo {author} {\bibfnamefont {B.}~\bibnamefont
  {Abelev}} \emph {et~al.} (\bibinfo {collaboration} {ALICE}),\ }\href
  {\doibase 10.1016/j.physletb.2013.01.012} {\bibfield  {journal} {\bibinfo
  {journal} {Phys. Lett. B}\ }\textbf {\bibinfo {volume} {719}},\ \bibinfo
  {pages} {29} (\bibinfo {year} {2013})},\ \Eprint
  {http://arxiv.org/abs/1212.2001} {arXiv:1212.2001 [nucl-ex]} \BibitemShut
  {NoStop}%
\bibitem [{\citenamefont {Khachatryan}\ \emph {et~al.}(2017)\citenamefont
  {Khachatryan} \emph {et~al.}}]{CMS:2016fnw}%
  \BibitemOpen
  \bibfield  {author} {\bibinfo {author} {\bibfnamefont {V.}~\bibnamefont
  {Khachatryan}} \emph {et~al.} (\bibinfo {collaboration} {CMS}),\ }\href
  {\doibase 10.1016/j.physletb.2016.12.009} {\bibfield  {journal} {\bibinfo
  {journal} {Phys. Lett. B}\ }\textbf {\bibinfo {volume} {765}},\ \bibinfo
  {pages} {193} (\bibinfo {year} {2017})},\ \Eprint
  {http://arxiv.org/abs/1606.06198} {arXiv:1606.06198 [nucl-ex]} \BibitemShut
  {NoStop}%
\bibitem [{\citenamefont {Schenke}(2021)}]{Schenke:2021mxx}%
  \BibitemOpen
  \bibfield  {author} {\bibinfo {author} {\bibfnamefont {B.}~\bibnamefont
  {Schenke}},\ }\href {\doibase 10.1088/1361-6633/ac14c9} {\bibfield  {journal}
  {\bibinfo  {journal} {Rept. Prog. Phys.}\ }\textbf {\bibinfo {volume} {84}},\
  \bibinfo {pages} {082301} (\bibinfo {year} {2021})},\ \Eprint
  {http://arxiv.org/abs/2102.11189} {arXiv:2102.11189 [nucl-th]} \BibitemShut
  {NoStop}%
\bibitem [{\citenamefont {Aoki}\ \emph {et~al.}(2006)\citenamefont {Aoki},
  \citenamefont {Endrodi}, \citenamefont {Fodor}, \citenamefont {Katz},\ and\
  \citenamefont {Szabo}}]{Aoki:2006we}%
  \BibitemOpen
  \bibfield  {author} {\bibinfo {author} {\bibfnamefont {Y.}~\bibnamefont
  {Aoki}}, \bibinfo {author} {\bibfnamefont {G.}~\bibnamefont {Endrodi}},
  \bibinfo {author} {\bibfnamefont {Z.}~\bibnamefont {Fodor}}, \bibinfo
  {author} {\bibfnamefont {S.~D.}\ \bibnamefont {Katz}}, \ and\ \bibinfo
  {author} {\bibfnamefont {K.~K.}\ \bibnamefont {Szabo}},\ }\href {\doibase
  10.1038/nature05120} {\bibfield  {journal} {\bibinfo  {journal} {Nature}\
  }\textbf {\bibinfo {volume} {443}},\ \bibinfo {pages} {675} (\bibinfo {year}
  {2006})},\ \Eprint {http://arxiv.org/abs/hep-lat/0611014}
  {arXiv:hep-lat/0611014} \BibitemShut {NoStop}%
\bibitem [{\citenamefont {Borsanyi}\ \emph {et~al.}(2020)\citenamefont
  {Borsanyi}, \citenamefont {Fodor}, \citenamefont {Guenther}, \citenamefont
  {Kara}, \citenamefont {Katz}, \citenamefont {Parotto}, \citenamefont
  {Pasztor}, \citenamefont {Ratti},\ and\ \citenamefont
  {Szabo}}]{Borsanyi:2020fev}%
  \BibitemOpen
  \bibfield  {author} {\bibinfo {author} {\bibfnamefont {S.}~\bibnamefont
  {Borsanyi}}, \bibinfo {author} {\bibfnamefont {Z.}~\bibnamefont {Fodor}},
  \bibinfo {author} {\bibfnamefont {J.~N.}\ \bibnamefont {Guenther}}, \bibinfo
  {author} {\bibfnamefont {R.}~\bibnamefont {Kara}}, \bibinfo {author}
  {\bibfnamefont {S.~D.}\ \bibnamefont {Katz}}, \bibinfo {author}
  {\bibfnamefont {P.}~\bibnamefont {Parotto}}, \bibinfo {author} {\bibfnamefont
  {A.}~\bibnamefont {Pasztor}}, \bibinfo {author} {\bibfnamefont
  {C.}~\bibnamefont {Ratti}}, \ and\ \bibinfo {author} {\bibfnamefont {K.~K.}\
  \bibnamefont {Szabo}},\ }\href {\doibase 10.1103/PhysRevLett.125.052001}
  {\bibfield  {journal} {\bibinfo  {journal} {Phys. Rev. Lett.}\ }\textbf
  {\bibinfo {volume} {125}},\ \bibinfo {pages} {052001} (\bibinfo {year}
  {2020})},\ \Eprint {http://arxiv.org/abs/2002.02821} {arXiv:2002.02821
  [hep-lat]} \BibitemShut {NoStop}%
\bibitem [{\citenamefont {Bazavov}\ \emph {et~al.}(2012)\citenamefont {Bazavov}
  \emph {et~al.}}]{Bazavov:2011nk}%
  \BibitemOpen
  \bibfield  {author} {\bibinfo {author} {\bibfnamefont {A.}~\bibnamefont
  {Bazavov}} \emph {et~al.},\ }\href {\doibase 10.1103/PhysRevD.85.054503}
  {\bibfield  {journal} {\bibinfo  {journal} {Phys. Rev. D}\ }\textbf {\bibinfo
  {volume} {85}},\ \bibinfo {pages} {054503} (\bibinfo {year} {2012})},\
  \Eprint {http://arxiv.org/abs/1111.1710} {arXiv:1111.1710 [hep-lat]}
  \BibitemShut {NoStop}%
\bibitem [{\citenamefont {Bazavov}\ \emph {et~al.}(2019)\citenamefont {Bazavov}
  \emph {et~al.}}]{HotQCD:2018pds}%
  \BibitemOpen
  \bibfield  {author} {\bibinfo {author} {\bibfnamefont {A.}~\bibnamefont
  {Bazavov}} \emph {et~al.} (\bibinfo {collaboration} {HotQCD}),\ }\href
  {\doibase 10.1016/j.physletb.2019.05.013} {\bibfield  {journal} {\bibinfo
  {journal} {Phys. Lett. B}\ }\textbf {\bibinfo {volume} {795}},\ \bibinfo
  {pages} {15} (\bibinfo {year} {2019})},\ \Eprint
  {http://arxiv.org/abs/1812.08235} {arXiv:1812.08235 [hep-lat]} \BibitemShut
  {NoStop}%
\bibitem [{\citenamefont {Guenther}(2021)}]{Guenther:2020jwe}%
  \BibitemOpen
  \bibfield  {author} {\bibinfo {author} {\bibfnamefont {J.~N.}\ \bibnamefont
  {Guenther}},\ }\href {\doibase 10.1140/epja/s10050-021-00354-6} {\bibfield
  {journal} {\bibinfo  {journal} {Eur. Phys. J. A}\ }\textbf {\bibinfo {volume}
  {57}},\ \bibinfo {pages} {136} (\bibinfo {year} {2021})},\ \Eprint
  {http://arxiv.org/abs/2010.15503} {arXiv:2010.15503 [hep-lat]} \BibitemShut
  {NoStop}%
\bibitem [{\citenamefont {Fodor}\ and\ \citenamefont
  {Katz}(2002{\natexlab{a}})}]{Fodor:2001au}%
  \BibitemOpen
  \bibfield  {author} {\bibinfo {author} {\bibfnamefont {Z.}~\bibnamefont
  {Fodor}}\ and\ \bibinfo {author} {\bibfnamefont {S.~D.}\ \bibnamefont
  {Katz}},\ }\href {\doibase 10.1016/S0370-2693(02)01583-6} {\bibfield
  {journal} {\bibinfo  {journal} {Phys. Lett. B}\ }\textbf {\bibinfo {volume}
  {534}},\ \bibinfo {pages} {87} (\bibinfo {year} {2002}{\natexlab{a}})},\
  \Eprint {http://arxiv.org/abs/hep-lat/0104001} {arXiv:hep-lat/0104001}
  \BibitemShut {NoStop}%
\bibitem [{\citenamefont {Fodor}\ and\ \citenamefont
  {Katz}(2002{\natexlab{b}})}]{Fodor:2001pe}%
  \BibitemOpen
  \bibfield  {author} {\bibinfo {author} {\bibfnamefont {Z.}~\bibnamefont
  {Fodor}}\ and\ \bibinfo {author} {\bibfnamefont {S.~D.}\ \bibnamefont
  {Katz}},\ }\href {\doibase 10.1088/1126-6708/2002/03/014} {\bibfield
  {journal} {\bibinfo  {journal} {JHEP}\ }\textbf {\bibinfo {volume} {03}},\
  \bibinfo {pages} {014} (\bibinfo {year} {2002}{\natexlab{b}})},\ \Eprint
  {http://arxiv.org/abs/hep-lat/0106002} {arXiv:hep-lat/0106002} \BibitemShut
  {NoStop}%
\bibitem [{\citenamefont {Csikor}\ \emph {et~al.}(2003)\citenamefont {Csikor},
  \citenamefont {Egri}, \citenamefont {Fodor}, \citenamefont {Katz},
  \citenamefont {Szabo},\ and\ \citenamefont {Toth}}]{Csikor:2002ic}%
  \BibitemOpen
  \bibfield  {author} {\bibinfo {author} {\bibfnamefont {F.}~\bibnamefont
  {Csikor}}, \bibinfo {author} {\bibfnamefont {G.~I.}\ \bibnamefont {Egri}},
  \bibinfo {author} {\bibfnamefont {Z.}~\bibnamefont {Fodor}}, \bibinfo
  {author} {\bibfnamefont {S.~D.}\ \bibnamefont {Katz}}, \bibinfo {author}
  {\bibfnamefont {K.~K.}\ \bibnamefont {Szabo}}, \ and\ \bibinfo {author}
  {\bibfnamefont {A.~I.}\ \bibnamefont {Toth}},\ }\href {\doibase
  10.1016/S0920-5632(03)80453-X} {\bibfield  {journal} {\bibinfo  {journal}
  {Nucl. Phys. B Proc. Suppl.}\ }\textbf {\bibinfo {volume} {119}},\ \bibinfo
  {pages} {547} (\bibinfo {year} {2003})},\ \Eprint
  {http://arxiv.org/abs/hep-lat/0209114} {arXiv:hep-lat/0209114} \BibitemShut
  {NoStop}%
\bibitem [{\citenamefont {Magdy}\ \emph {et~al.}(2022)\citenamefont {Magdy},
  \citenamefont {Parfenov}, \citenamefont {Taranenko}, \citenamefont
  {Karpenko},\ and\ \citenamefont {Lacey}}]{Magdy:2021cci}%
  \BibitemOpen
  \bibfield  {author} {\bibinfo {author} {\bibfnamefont {N.}~\bibnamefont
  {Magdy}}, \bibinfo {author} {\bibfnamefont {P.}~\bibnamefont {Parfenov}},
  \bibinfo {author} {\bibfnamefont {A.}~\bibnamefont {Taranenko}}, \bibinfo
  {author} {\bibfnamefont {I.}~\bibnamefont {Karpenko}}, \ and\ \bibinfo
  {author} {\bibfnamefont {R.~A.}\ \bibnamefont {Lacey}},\ }\href {\doibase
  10.1103/PhysRevC.105.044901} {\bibfield  {journal} {\bibinfo  {journal}
  {Phys. Rev. C}\ }\textbf {\bibinfo {volume} {105}},\ \bibinfo {pages}
  {044901} (\bibinfo {year} {2022})},\ \Eprint
  {http://arxiv.org/abs/2111.07406} {arXiv:2111.07406 [nucl-th]} \BibitemShut
  {NoStop}%
\bibitem [{\citenamefont {Adamczyk}\ \emph {et~al.}(2012)\citenamefont
  {Adamczyk} \emph {et~al.}}]{STAR:2012och}%
  \BibitemOpen
  \bibfield  {author} {\bibinfo {author} {\bibfnamefont {L.}~\bibnamefont
  {Adamczyk}} \emph {et~al.} (\bibinfo {collaboration} {STAR}),\ }\href
  {\doibase 10.1103/PhysRevC.86.054908} {\bibfield  {journal} {\bibinfo
  {journal} {Phys. Rev. C}\ }\textbf {\bibinfo {volume} {86}},\ \bibinfo
  {pages} {054908} (\bibinfo {year} {2012})}\BibitemShut {NoStop}%
\bibitem [{\citenamefont {Becattini}\ \emph {et~al.}(2006)\citenamefont
  {Becattini}, \citenamefont {Manninen},\ and\ \citenamefont
  {Gazdzicki}}]{Becattini:2005xt}%
  \BibitemOpen
  \bibfield  {author} {\bibinfo {author} {\bibfnamefont {F.}~\bibnamefont
  {Becattini}}, \bibinfo {author} {\bibfnamefont {J.}~\bibnamefont {Manninen}},
  \ and\ \bibinfo {author} {\bibfnamefont {M.}~\bibnamefont {Gazdzicki}},\
  }\href {\doibase 10.1103/PhysRevC.73.044905} {\bibfield  {journal} {\bibinfo
  {journal} {Phys. Rev. C}\ }\textbf {\bibinfo {volume} {73}},\ \bibinfo
  {pages} {044905} (\bibinfo {year} {2006})},\ \Eprint
  {http://arxiv.org/abs/hep-ph/0511092} {arXiv:hep-ph/0511092} \BibitemShut
  {NoStop}%
\bibitem [{\citenamefont {Werner}\ \emph {et~al.}(2024)\citenamefont {Werner},
  \citenamefont {Jahan}, \citenamefont {Karpenko}, \citenamefont {Pierog},
  \citenamefont {Stefaniak},\ and\ \citenamefont {Vintache}}]{Werner:2024ntd}%
  \BibitemOpen
  \bibfield  {author} {\bibinfo {author} {\bibfnamefont {K.}~\bibnamefont
  {Werner}}, \bibinfo {author} {\bibfnamefont {J.}~\bibnamefont {Jahan}},
  \bibinfo {author} {\bibfnamefont {I.}~\bibnamefont {Karpenko}}, \bibinfo
  {author} {\bibfnamefont {T.}~\bibnamefont {Pierog}}, \bibinfo {author}
  {\bibfnamefont {M.}~\bibnamefont {Stefaniak}}, \ and\ \bibinfo {author}
  {\bibfnamefont {D.}~\bibnamefont {Vintache}},\ }\href@noop {} {\  (\bibinfo
  {year} {2024})},\ \Eprint {http://arxiv.org/abs/2401.11275} {arXiv:2401.11275
  [hep-ph]} \BibitemShut {NoStop}%
\bibitem [{\citenamefont {Cimerman}\ \emph {et~al.}(2023)\citenamefont
  {Cimerman}, \citenamefont {Karpenko}, \citenamefont {Tomasik},\ and\
  \citenamefont {Huovinen}}]{Cimerman:2023hjw}%
  \BibitemOpen
  \bibfield  {author} {\bibinfo {author} {\bibfnamefont {J.}~\bibnamefont
  {Cimerman}}, \bibinfo {author} {\bibfnamefont {I.}~\bibnamefont {Karpenko}},
  \bibinfo {author} {\bibfnamefont {B.}~\bibnamefont {Tomasik}}, \ and\
  \bibinfo {author} {\bibfnamefont {P.}~\bibnamefont {Huovinen}},\ }\href
  {\doibase 10.1103/PhysRevC.107.044902} {\bibfield  {journal} {\bibinfo
  {journal} {Phys. Rev. C}\ }\textbf {\bibinfo {volume} {107}},\ \bibinfo
  {pages} {044902} (\bibinfo {year} {2023})},\ \Eprint
  {http://arxiv.org/abs/2301.11894} {arXiv:2301.11894 [nucl-th]} \BibitemShut
  {NoStop}%
\bibitem [{\citenamefont {Aboona}\ \emph {et~al.}(2023)\citenamefont {Aboona}
  \emph {et~al.}}]{STAR:2022vkx}%
  \BibitemOpen
  \bibfield  {author} {\bibinfo {author} {\bibfnamefont {B.}~\bibnamefont
  {Aboona}} \emph {et~al.} (\bibinfo {collaboration} {STAR}),\ }\href {\doibase
  10.1016/j.physletb.2023.137755} {\bibfield  {journal} {\bibinfo  {journal}
  {Phys. Lett. B}\ }\textbf {\bibinfo {volume} {839}},\ \bibinfo {pages}
  {137755} (\bibinfo {year} {2023})},\ \Eprint
  {http://arxiv.org/abs/2211.11637} {arXiv:2211.11637 [nucl-ex]} \BibitemShut
  {NoStop}%
\bibitem [{\citenamefont {Abdallah}\ \emph {et~al.}(2022)\citenamefont
  {Abdallah} \emph {et~al.}}]{STAR:2022gki}%
  \BibitemOpen
  \bibfield  {author} {\bibinfo {author} {\bibfnamefont {M.}~\bibnamefont
  {Abdallah}} \emph {et~al.} (\bibinfo {collaboration} {STAR}),\ }\href
  {\doibase 10.1103/PhysRevLett.129.252301} {\bibfield  {journal} {\bibinfo
  {journal} {Phys. Rev. Lett.}\ }\textbf {\bibinfo {volume} {129}},\ \bibinfo
  {pages} {252301} (\bibinfo {year} {2022})},\ \Eprint
  {http://arxiv.org/abs/2201.10365} {arXiv:2201.10365 [nucl-ex]} \BibitemShut
  {NoStop}%
\bibitem [{\citenamefont {Rao}\ \emph {et~al.}(2021)\citenamefont {Rao},
  \citenamefont {Sievert},\ and\ \citenamefont
  {Noronha-Hostler}}]{Rao:2019vgy}%
  \BibitemOpen
  \bibfield  {author} {\bibinfo {author} {\bibfnamefont {S.}~\bibnamefont
  {Rao}}, \bibinfo {author} {\bibfnamefont {M.}~\bibnamefont {Sievert}}, \ and\
  \bibinfo {author} {\bibfnamefont {J.}~\bibnamefont {Noronha-Hostler}},\
  }\href {\doibase 10.1103/PhysRevC.103.034910} {\bibfield  {journal} {\bibinfo
   {journal} {Phys. Rev. C}\ }\textbf {\bibinfo {volume} {103}},\ \bibinfo
  {pages} {034910} (\bibinfo {year} {2021})},\ \Eprint
  {http://arxiv.org/abs/1910.03677} {arXiv:1910.03677 [nucl-th]} \BibitemShut
  {NoStop}%
\bibitem [{\citenamefont {Adamczyk}\ \emph {et~al.}(2018)\citenamefont
  {Adamczyk} \emph {et~al.}}]{STAR:2017ieb}%
  \BibitemOpen
  \bibfield  {author} {\bibinfo {author} {\bibfnamefont {L.}~\bibnamefont
  {Adamczyk}} \emph {et~al.} (\bibinfo {collaboration} {STAR}),\ }\href
  {\doibase 10.1103/PhysRevLett.121.032301} {\bibfield  {journal} {\bibinfo
  {journal} {Phys. Rev. Lett.}\ }\textbf {\bibinfo {volume} {121}},\ \bibinfo
  {pages} {032301} (\bibinfo {year} {2018})},\ \Eprint
  {http://arxiv.org/abs/1707.01988} {arXiv:1707.01988 [nucl-ex]} \BibitemShut
  {NoStop}%
\bibitem [{\citenamefont {Buckley}\ \emph {et~al.}(2013)\citenamefont
  {Buckley}, \citenamefont {Butterworth}, \citenamefont {Grellscheid},
  \citenamefont {Hoeth}, \citenamefont {Lonnblad}, \citenamefont {Monk},
  \citenamefont {Schulz},\ and\ \citenamefont {Siegert}}]{Buckley:2010ar}%
  \BibitemOpen
  \bibfield  {author} {\bibinfo {author} {\bibfnamefont {A.}~\bibnamefont
  {Buckley}}, \bibinfo {author} {\bibfnamefont {J.}~\bibnamefont
  {Butterworth}}, \bibinfo {author} {\bibfnamefont {D.}~\bibnamefont
  {Grellscheid}}, \bibinfo {author} {\bibfnamefont {H.}~\bibnamefont {Hoeth}},
  \bibinfo {author} {\bibfnamefont {L.}~\bibnamefont {Lonnblad}}, \bibinfo
  {author} {\bibfnamefont {J.}~\bibnamefont {Monk}}, \bibinfo {author}
  {\bibfnamefont {H.}~\bibnamefont {Schulz}}, \ and\ \bibinfo {author}
  {\bibfnamefont {F.}~\bibnamefont {Siegert}},\ }\href {\doibase
  10.1016/j.cpc.2013.05.021} {\bibfield  {journal} {\bibinfo  {journal}
  {Comput. Phys. Commun.}\ }\textbf {\bibinfo {volume} {184}},\ \bibinfo
  {pages} {2803} (\bibinfo {year} {2013})},\ \Eprint
  {http://arxiv.org/abs/1003.0694} {arXiv:1003.0694 [hep-ph]} \BibitemShut
  {NoStop}%
\bibitem [{\citenamefont {Bierlich}\ \emph
  {et~al.}(2020{\natexlab{a}})\citenamefont {Bierlich} \emph
  {et~al.}}]{Bierlich:2019rhm}%
  \BibitemOpen
  \bibfield  {author} {\bibinfo {author} {\bibfnamefont {C.}~\bibnamefont
  {Bierlich}} \emph {et~al.},\ }\href {\doibase 10.21468/SciPostPhys.8.2.026}
  {\bibfield  {journal} {\bibinfo  {journal} {SciPost Phys.}\ }\textbf
  {\bibinfo {volume} {8}},\ \bibinfo {pages} {026} (\bibinfo {year}
  {2020}{\natexlab{a}})},\ \Eprint {http://arxiv.org/abs/1912.05451}
  {arXiv:1912.05451 [hep-ph]} \BibitemShut {NoStop}%
\bibitem [{\citenamefont {Adcox}\ \emph {et~al.}(2001)\citenamefont {Adcox}
  \emph {et~al.}}]{PHENIX:2001kdi}%
  \BibitemOpen
  \bibfield  {author} {\bibinfo {author} {\bibfnamefont {K.}~\bibnamefont
  {Adcox}} \emph {et~al.} (\bibinfo {collaboration} {PHENIX}),\ }\href
  {\doibase 10.1103/PhysRevLett.87.052301} {\bibfield  {journal} {\bibinfo
  {journal} {Phys. Rev. Lett.}\ }\textbf {\bibinfo {volume} {87}},\ \bibinfo
  {pages} {052301} (\bibinfo {year} {2001})},\ \Eprint
  {http://arxiv.org/abs/nucl-ex/0104015} {arXiv:nucl-ex/0104015} \BibitemShut
  {NoStop}%
\bibitem [{\citenamefont {Adler}\ \emph {et~al.}(2014)\citenamefont {Adler}
  \emph {et~al.}}]{PHENIX:2013ehw}%
  \BibitemOpen
  \bibfield  {author} {\bibinfo {author} {\bibfnamefont {S.~S.}\ \bibnamefont
  {Adler}} \emph {et~al.} (\bibinfo {collaboration} {PHENIX}),\ }\href
  {\doibase 10.1103/PhysRevC.89.044905} {\bibfield  {journal} {\bibinfo
  {journal} {Phys. Rev. C}\ }\textbf {\bibinfo {volume} {89}},\ \bibinfo
  {pages} {044905} (\bibinfo {year} {2014})},\ \Eprint
  {http://arxiv.org/abs/1312.6676} {arXiv:1312.6676 [nucl-ex]} \BibitemShut
  {NoStop}%
\bibitem [{\citenamefont {Bierlich}\ \emph {et~al.}(2018)\citenamefont
  {Bierlich}, \citenamefont {Gustafson}, \citenamefont {L\"onnblad},\ and\
  \citenamefont {Shah}}]{Bierlich:2018xfw}%
  \BibitemOpen
  \bibfield  {author} {\bibinfo {author} {\bibfnamefont {C.}~\bibnamefont
  {Bierlich}}, \bibinfo {author} {\bibfnamefont {G.}~\bibnamefont {Gustafson}},
  \bibinfo {author} {\bibfnamefont {L.}~\bibnamefont {L\"onnblad}}, \ and\
  \bibinfo {author} {\bibfnamefont {H.}~\bibnamefont {Shah}},\ }\href {\doibase
  10.1007/JHEP10(2018)134} {\bibfield  {journal} {\bibinfo  {journal} {JHEP}\
  }\textbf {\bibinfo {volume} {10}},\ \bibinfo {pages} {134} (\bibinfo {year}
  {2018})},\ \Eprint {http://arxiv.org/abs/1806.10820} {arXiv:1806.10820
  [hep-ph]} \BibitemShut {NoStop}%
\bibitem [{\citenamefont {Lin}\ \emph {et~al.}(2005)\citenamefont {Lin},
  \citenamefont {Ko}, \citenamefont {Li}, \citenamefont {Zhang},\ and\
  \citenamefont {Pal}}]{Lin:2004en}%
  \BibitemOpen
  \bibfield  {author} {\bibinfo {author} {\bibfnamefont {Z.-W.}\ \bibnamefont
  {Lin}}, \bibinfo {author} {\bibfnamefont {C.~M.}\ \bibnamefont {Ko}},
  \bibinfo {author} {\bibfnamefont {B.-A.}\ \bibnamefont {Li}}, \bibinfo
  {author} {\bibfnamefont {B.}~\bibnamefont {Zhang}}, \ and\ \bibinfo {author}
  {\bibfnamefont {S.}~\bibnamefont {Pal}},\ }\href {\doibase
  10.1103/PhysRevC.72.064901} {\bibfield  {journal} {\bibinfo  {journal} {Phys.
  Rev.}\ }\textbf {\bibinfo {volume} {C72}},\ \bibinfo {pages} {064901}
  (\bibinfo {year} {2005})},\ \Eprint {http://arxiv.org/abs/nucl-th/0411110}
  {arXiv:nucl-th/0411110 [nucl-th]} \BibitemShut {NoStop}%
\bibitem [{\citenamefont {Wang}\ and\ \citenamefont
  {Gyulassy}(1991)}]{Wang:1991hta}%
  \BibitemOpen
  \bibfield  {author} {\bibinfo {author} {\bibfnamefont {X.-N.}\ \bibnamefont
  {Wang}}\ and\ \bibinfo {author} {\bibfnamefont {M.}~\bibnamefont
  {Gyulassy}},\ }\href {\doibase 10.1103/PhysRevD.44.3501} {\bibfield
  {journal} {\bibinfo  {journal} {Phys. Rev.}\ }\textbf {\bibinfo {volume}
  {D44}},\ \bibinfo {pages} {3501} (\bibinfo {year} {1991})}\BibitemShut
  {NoStop}%
\bibitem [{\citenamefont {Gyulassy}\ and\ \citenamefont
  {Wang}(1994)}]{Gyulassy:1994ew}%
  \BibitemOpen
  \bibfield  {author} {\bibinfo {author} {\bibfnamefont {M.}~\bibnamefont
  {Gyulassy}}\ and\ \bibinfo {author} {\bibfnamefont {X.-N.}\ \bibnamefont
  {Wang}},\ }\href {\doibase 10.1016/0010-4655(94)90057-4} {\bibfield
  {journal} {\bibinfo  {journal} {Comput. Phys. Commun.}\ }\textbf {\bibinfo
  {volume} {83}},\ \bibinfo {pages} {307} (\bibinfo {year} {1994})},\ \Eprint
  {http://arxiv.org/abs/nucl-th/9502021} {arXiv:nucl-th/9502021 [nucl-th]}
  \BibitemShut {NoStop}%
\bibitem [{\citenamefont {Weil}\ \emph {et~al.}(2016)\citenamefont {Weil} \emph
  {et~al.}}]{SMASH:2016zqf}%
  \BibitemOpen
  \bibfield  {author} {\bibinfo {author} {\bibfnamefont {J.}~\bibnamefont
  {Weil}} \emph {et~al.} (\bibinfo {collaboration} {SMASH}),\ }\href {\doibase
  10.1103/PhysRevC.94.054905} {\bibfield  {journal} {\bibinfo  {journal} {Phys.
  Rev. C}\ }\textbf {\bibinfo {volume} {94}},\ \bibinfo {pages} {054905}
  (\bibinfo {year} {2016})},\ \Eprint {http://arxiv.org/abs/1606.06642}
  {arXiv:1606.06642 [nucl-th]} \BibitemShut {NoStop}%
\bibitem [{\citenamefont {Glauber}(1955)}]{Glauber:1955qq}%
  \BibitemOpen
  \bibfield  {author} {\bibinfo {author} {\bibfnamefont {R.~J.}\ \bibnamefont
  {Glauber}},\ }\href {\doibase 10.1103/PhysRev.100.242} {\bibfield  {journal}
  {\bibinfo  {journal} {Phys. Rev.}\ }\textbf {\bibinfo {volume} {100}},\
  \bibinfo {pages} {242} (\bibinfo {year} {1955})}\BibitemShut {NoStop}%
\bibitem [{\citenamefont {Rybczynski}\ \emph {et~al.}(2014)\citenamefont
  {Rybczynski}, \citenamefont {Stefanek}, \citenamefont {Broniowski},\ and\
  \citenamefont {Bozek}}]{Rybczynski:2013yba}%
  \BibitemOpen
  \bibfield  {author} {\bibinfo {author} {\bibfnamefont {M.}~\bibnamefont
  {Rybczynski}}, \bibinfo {author} {\bibfnamefont {G.}~\bibnamefont
  {Stefanek}}, \bibinfo {author} {\bibfnamefont {W.}~\bibnamefont
  {Broniowski}}, \ and\ \bibinfo {author} {\bibfnamefont {P.}~\bibnamefont
  {Bozek}},\ }\href {\doibase 10.1016/j.cpc.2014.02.016} {\bibfield  {journal}
  {\bibinfo  {journal} {Comput. Phys. Commun.}\ }\textbf {\bibinfo {volume}
  {185}},\ \bibinfo {pages} {1759} (\bibinfo {year} {2014})},\ \Eprint
  {http://arxiv.org/abs/1310.5475} {arXiv:1310.5475 [nucl-th]} \BibitemShut
  {NoStop}%
\bibitem [{\citenamefont {Sjostrand}\ and\ \citenamefont
  {Bengtsson}(1987)}]{Sjostrand:1986hx}%
  \BibitemOpen
  \bibfield  {author} {\bibinfo {author} {\bibfnamefont {T.}~\bibnamefont
  {Sjostrand}}\ and\ \bibinfo {author} {\bibfnamefont {M.}~\bibnamefont
  {Bengtsson}},\ }\href {\doibase 10.1016/0010-4655(87)90054-3} {\bibfield
  {journal} {\bibinfo  {journal} {Comput. Phys. Commun.}\ }\textbf {\bibinfo
  {volume} {43}},\ \bibinfo {pages} {367} (\bibinfo {year} {1987})}\BibitemShut
  {NoStop}%
\bibitem [{\citenamefont {Andersson}\ \emph {et~al.}(1983)\citenamefont
  {Andersson}, \citenamefont {Gustafson}, \citenamefont {Ingelman},\ and\
  \citenamefont {Sjostrand}}]{Andersson:1983ia}%
  \BibitemOpen
  \bibfield  {author} {\bibinfo {author} {\bibfnamefont {B.}~\bibnamefont
  {Andersson}}, \bibinfo {author} {\bibfnamefont {G.}~\bibnamefont
  {Gustafson}}, \bibinfo {author} {\bibfnamefont {G.}~\bibnamefont {Ingelman}},
  \ and\ \bibinfo {author} {\bibfnamefont {T.}~\bibnamefont {Sjostrand}},\
  }\href {\doibase 10.1016/0370-1573(83)90080-7} {\bibfield  {journal}
  {\bibinfo  {journal} {Phys. Rept.}\ }\textbf {\bibinfo {volume} {97}},\
  \bibinfo {pages} {31} (\bibinfo {year} {1983})}\BibitemShut {NoStop}%
\bibitem [{\citenamefont {Ma}\ and\ \citenamefont {Lin}(2016)}]{Ma:2016fve}%
  \BibitemOpen
  \bibfield  {author} {\bibinfo {author} {\bibfnamefont {G.-L.}\ \bibnamefont
  {Ma}}\ and\ \bibinfo {author} {\bibfnamefont {Z.-W.}\ \bibnamefont {Lin}},\
  }\href {\doibase 10.1103/PhysRevC.93.054911} {\bibfield  {journal} {\bibinfo
  {journal} {Phys. Rev.}\ }\textbf {\bibinfo {volume} {C93}},\ \bibinfo {pages}
  {054911} (\bibinfo {year} {2016})},\ \Eprint
  {http://arxiv.org/abs/1601.08160} {arXiv:1601.08160 [nucl-th]} \BibitemShut
  {NoStop}%
\bibitem [{\citenamefont {Ma}(2013)}]{Ma:2013gga}%
  \BibitemOpen
  \bibfield  {author} {\bibinfo {author} {\bibfnamefont {G.-L.}\ \bibnamefont
  {Ma}},\ }\href {\doibase 10.1103/PhysRevC.88.021902} {\bibfield  {journal}
  {\bibinfo  {journal} {Phys. Rev.}\ }\textbf {\bibinfo {volume} {C88}},\
  \bibinfo {pages} {021902} (\bibinfo {year} {2013})},\ \Eprint
  {http://arxiv.org/abs/1306.1306} {arXiv:1306.1306 [nucl-th]} \BibitemShut
  {NoStop}%
\bibitem [{\citenamefont {Ma}(2014)}]{Ma:2013uqa}%
  \BibitemOpen
  \bibfield  {author} {\bibinfo {author} {\bibfnamefont {G.-L.}\ \bibnamefont
  {Ma}},\ }\href {\doibase 10.1103/PhysRevC.89.024902} {\bibfield  {journal}
  {\bibinfo  {journal} {Phys. Rev.}\ }\textbf {\bibinfo {volume} {C89}},\
  \bibinfo {pages} {024902} (\bibinfo {year} {2014})},\ \Eprint
  {http://arxiv.org/abs/1309.5555} {arXiv:1309.5555 [nucl-th]} \BibitemShut
  {NoStop}%
\bibitem [{\citenamefont {Bzdak}\ and\ \citenamefont
  {Ma}(2014)}]{Bzdak:2014dia}%
  \BibitemOpen
  \bibfield  {author} {\bibinfo {author} {\bibfnamefont {A.}~\bibnamefont
  {Bzdak}}\ and\ \bibinfo {author} {\bibfnamefont {G.-L.}\ \bibnamefont {Ma}},\
  }\href {\doibase 10.1103/PhysRevLett.113.252301} {\bibfield  {journal}
  {\bibinfo  {journal} {Phys. Rev. Lett.}\ }\textbf {\bibinfo {volume} {113}},\
  \bibinfo {pages} {252301} (\bibinfo {year} {2014})},\ \Eprint
  {http://arxiv.org/abs/1406.2804} {arXiv:1406.2804 [hep-ph]} \BibitemShut
  {NoStop}%
\bibitem [{\citenamefont {Nie}\ \emph {et~al.}(2018)\citenamefont {Nie},
  \citenamefont {Huo}, \citenamefont {Jia},\ and\ \citenamefont
  {Ma}}]{Nie:2018xog}%
  \BibitemOpen
  \bibfield  {author} {\bibinfo {author} {\bibfnamefont {M.-W.}\ \bibnamefont
  {Nie}}, \bibinfo {author} {\bibfnamefont {P.}~\bibnamefont {Huo}}, \bibinfo
  {author} {\bibfnamefont {J.}~\bibnamefont {Jia}}, \ and\ \bibinfo {author}
  {\bibfnamefont {G.-L.}\ \bibnamefont {Ma}},\ }\href {\doibase
  10.1103/PhysRevC.98.034903} {\bibfield  {journal} {\bibinfo  {journal} {Phys.
  Rev.}\ }\textbf {\bibinfo {volume} {C98}},\ \bibinfo {pages} {034903}
  (\bibinfo {year} {2018})},\ \Eprint {http://arxiv.org/abs/1802.00374}
  {arXiv:1802.00374 [hep-ph]} \BibitemShut {NoStop}%
\bibitem [{\citenamefont {Haque}\ \emph {et~al.}(2019)\citenamefont {Haque},
  \citenamefont {Nasim},\ and\ \citenamefont {Mohanty}}]{Haque:2019vgi}%
  \BibitemOpen
  \bibfield  {author} {\bibinfo {author} {\bibfnamefont {M.~R.}\ \bibnamefont
  {Haque}}, \bibinfo {author} {\bibfnamefont {M.}~\bibnamefont {Nasim}}, \ and\
  \bibinfo {author} {\bibfnamefont {B.}~\bibnamefont {Mohanty}},\ }\href
  {\doibase 10.1088/1361-6471/ab2ba4} {\bibfield  {journal} {\bibinfo
  {journal} {J. Phys. G}\ }\textbf {\bibinfo {volume} {46}},\ \bibinfo {pages}
  {085104} (\bibinfo {year} {2019})}\BibitemShut {NoStop}%
\bibitem [{\citenamefont {Zhao}\ \emph {et~al.}(2020)\citenamefont {Zhao},
  \citenamefont {Feng}, \citenamefont {Li},\ and\ \citenamefont
  {Wang}}]{Zhao:2019kyk}%
  \BibitemOpen
  \bibfield  {author} {\bibinfo {author} {\bibfnamefont {J.}~\bibnamefont
  {Zhao}}, \bibinfo {author} {\bibfnamefont {Y.}~\bibnamefont {Feng}}, \bibinfo
  {author} {\bibfnamefont {H.}~\bibnamefont {Li}}, \ and\ \bibinfo {author}
  {\bibfnamefont {F.}~\bibnamefont {Wang}},\ }\href {\doibase
  10.1103/PhysRevC.101.034912} {\bibfield  {journal} {\bibinfo  {journal}
  {Phys. Rev. C}\ }\textbf {\bibinfo {volume} {101}},\ \bibinfo {pages}
  {034912} (\bibinfo {year} {2020})},\ \Eprint
  {http://arxiv.org/abs/1912.00299} {arXiv:1912.00299 [nucl-th]} \BibitemShut
  {NoStop}%
\bibitem [{\citenamefont {Bhaduri}\ and\ \citenamefont
  {Chattopadhyay}(2010)}]{Bhaduri:2010wi}%
  \BibitemOpen
  \bibfield  {author} {\bibinfo {author} {\bibfnamefont {P.~P.}\ \bibnamefont
  {Bhaduri}}\ and\ \bibinfo {author} {\bibfnamefont {S.}~\bibnamefont
  {Chattopadhyay}},\ }\href {\doibase 10.1103/PhysRevC.81.034906} {\bibfield
  {journal} {\bibinfo  {journal} {Phys. Rev. C}\ }\textbf {\bibinfo {volume}
  {81}},\ \bibinfo {pages} {034906} (\bibinfo {year} {2010})},\ \Eprint
  {http://arxiv.org/abs/1002.4100} {arXiv:1002.4100 [hep-ph]} \BibitemShut
  {NoStop}%
\bibitem [{\citenamefont {Nasim}\ \emph {et~al.}(2010)\citenamefont {Nasim},
  \citenamefont {Kumar}, \citenamefont {Netrakanti},\ and\ \citenamefont
  {Mohanty}}]{Nasim:2010hw}%
  \BibitemOpen
  \bibfield  {author} {\bibinfo {author} {\bibfnamefont {M.}~\bibnamefont
  {Nasim}}, \bibinfo {author} {\bibfnamefont {L.}~\bibnamefont {Kumar}},
  \bibinfo {author} {\bibfnamefont {P.~K.}\ \bibnamefont {Netrakanti}}, \ and\
  \bibinfo {author} {\bibfnamefont {B.}~\bibnamefont {Mohanty}},\ }\href
  {\doibase 10.1103/PhysRevC.82.054908} {\bibfield  {journal} {\bibinfo
  {journal} {Phys. Rev. C}\ }\textbf {\bibinfo {volume} {82}},\ \bibinfo
  {pages} {054908} (\bibinfo {year} {2010})},\ \Eprint
  {http://arxiv.org/abs/1010.5196} {arXiv:1010.5196 [nucl-ex]} \BibitemShut
  {NoStop}%
\bibitem [{\citenamefont {Xu}\ and\ \citenamefont
  {Ko}(2011{\natexlab{a}})}]{Xu:2010du}%
  \BibitemOpen
  \bibfield  {author} {\bibinfo {author} {\bibfnamefont {J.}~\bibnamefont
  {Xu}}\ and\ \bibinfo {author} {\bibfnamefont {C.~M.}\ \bibnamefont {Ko}},\
  }\href {\doibase 10.1103/PhysRevC.83.021903} {\bibfield  {journal} {\bibinfo
  {journal} {Phys. Rev. C}\ }\textbf {\bibinfo {volume} {83}},\ \bibinfo
  {pages} {021903} (\bibinfo {year} {2011}{\natexlab{a}})},\ \Eprint
  {http://arxiv.org/abs/1011.3750} {arXiv:1011.3750 [nucl-th]} \BibitemShut
  {NoStop}%
\bibitem [{\citenamefont {Magdy}\ \emph
  {et~al.}(2020{\natexlab{a}})\citenamefont {Magdy}, \citenamefont
  {Evdokimov},\ and\ \citenamefont {Lacey}}]{Magdy:2020bhd}%
  \BibitemOpen
  \bibfield  {author} {\bibinfo {author} {\bibfnamefont {N.}~\bibnamefont
  {Magdy}}, \bibinfo {author} {\bibfnamefont {O.}~\bibnamefont {Evdokimov}}, \
  and\ \bibinfo {author} {\bibfnamefont {R.~A.}\ \bibnamefont {Lacey}},\ }\href
  {\doibase 10.1088/1361-6471/abcb59} {\bibfield  {journal} {\bibinfo
  {journal} {J. Phys. G}\ }\textbf {\bibinfo {volume} {48}},\ \bibinfo {pages}
  {025101} (\bibinfo {year} {2020}{\natexlab{a}})},\ \Eprint
  {http://arxiv.org/abs/2002.04583} {arXiv:2002.04583 [nucl-ex]} \BibitemShut
  {NoStop}%
\bibitem [{\citenamefont {Guo}\ \emph {et~al.}(2019)\citenamefont {Guo},
  \citenamefont {Shi}, \citenamefont {Feng},\ and\ \citenamefont
  {Liao}}]{Guo:2019joy}%
  \BibitemOpen
  \bibfield  {author} {\bibinfo {author} {\bibfnamefont {Y.}~\bibnamefont
  {Guo}}, \bibinfo {author} {\bibfnamefont {S.}~\bibnamefont {Shi}}, \bibinfo
  {author} {\bibfnamefont {S.}~\bibnamefont {Feng}}, \ and\ \bibinfo {author}
  {\bibfnamefont {J.}~\bibnamefont {Liao}},\ }\href {\doibase
  10.1016/j.physletb.2019.134929} {\bibfield  {journal} {\bibinfo  {journal}
  {Phys. Lett. B}\ }\textbf {\bibinfo {volume} {798}},\ \bibinfo {pages}
  {134929} (\bibinfo {year} {2019})},\ \Eprint
  {http://arxiv.org/abs/1905.12613} {arXiv:1905.12613 [nucl-th]} \BibitemShut
  {NoStop}%
\bibitem [{\citenamefont {Magdy}\ \emph
  {et~al.}(2020{\natexlab{b}})\citenamefont {Magdy}, \citenamefont {Sun},
  \citenamefont {Ye}, \citenamefont {Evdokimov},\ and\ \citenamefont
  {Lacey}}]{Magdy:2020gxf}%
  \BibitemOpen
  \bibfield  {author} {\bibinfo {author} {\bibfnamefont {N.}~\bibnamefont
  {Magdy}}, \bibinfo {author} {\bibfnamefont {X.}~\bibnamefont {Sun}}, \bibinfo
  {author} {\bibfnamefont {Z.}~\bibnamefont {Ye}}, \bibinfo {author}
  {\bibfnamefont {O.}~\bibnamefont {Evdokimov}}, \ and\ \bibinfo {author}
  {\bibfnamefont {R.}~\bibnamefont {Lacey}},\ }\href {\doibase
  10.3390/universe6090146} {\bibfield  {journal} {\bibinfo  {journal}
  {Universe}\ }\textbf {\bibinfo {volume} {6}},\ \bibinfo {pages} {146}
  (\bibinfo {year} {2020}{\natexlab{b}})},\ \Eprint
  {http://arxiv.org/abs/2009.02734} {arXiv:2009.02734 [nucl-ex]} \BibitemShut
  {NoStop}%
\bibitem [{\citenamefont {Magdy}(2022)}]{Magdy:2022cvt}%
  \BibitemOpen
  \bibfield  {author} {\bibinfo {author} {\bibfnamefont {N.}~\bibnamefont
  {Magdy}},\ }\href@noop {} {\  (\bibinfo {year} {2022})},\ \Eprint
  {http://arxiv.org/abs/2206.05332} {arXiv:2206.05332 [nucl-th]} \BibitemShut
  {NoStop}%
\bibitem [{\citenamefont {Ferreres-Sol\'e}\ and\ \citenamefont
  {Sj\"ostrand}(2018)}]{Ferreres-Sole:2018vgo}%
  \BibitemOpen
  \bibfield  {author} {\bibinfo {author} {\bibfnamefont {S.}~\bibnamefont
  {Ferreres-Sol\'e}}\ and\ \bibinfo {author} {\bibfnamefont {T.}~\bibnamefont
  {Sj\"ostrand}},\ }\href {\doibase 10.1140/epjc/s10052-018-6459-8} {\bibfield
  {journal} {\bibinfo  {journal} {Eur. Phys. J. C}\ }\textbf {\bibinfo {volume}
  {78}},\ \bibinfo {pages} {983} (\bibinfo {year} {2018})},\ \Eprint
  {http://arxiv.org/abs/1808.04619} {arXiv:1808.04619 [hep-ph]} \BibitemShut
  {NoStop}%
\bibitem [{\citenamefont {Xu}\ and\ \citenamefont
  {Ko}(2011{\natexlab{b}})}]{Xu:2011fi}%
  \BibitemOpen
  \bibfield  {author} {\bibinfo {author} {\bibfnamefont {J.}~\bibnamefont
  {Xu}}\ and\ \bibinfo {author} {\bibfnamefont {C.~M.}\ \bibnamefont {Ko}},\
  }\href {\doibase 10.1103/PhysRevC.83.034904} {\bibfield  {journal} {\bibinfo
  {journal} {Phys. Rev. C}\ }\textbf {\bibinfo {volume} {83}},\ \bibinfo
  {pages} {034904} (\bibinfo {year} {2011}{\natexlab{b}})},\ \Eprint
  {http://arxiv.org/abs/1101.2231} {arXiv:1101.2231 [nucl-th]} \BibitemShut
  {NoStop}%
\bibitem [{\citenamefont {Zhang}(1998)}]{Zhang:1997ej}%
  \BibitemOpen
  \bibfield  {author} {\bibinfo {author} {\bibfnamefont {B.}~\bibnamefont
  {Zhang}},\ }\href {\doibase 10.1016/S0010-4655(98)00010-1} {\bibfield
  {journal} {\bibinfo  {journal} {Comput. Phys. Commun.}\ }\textbf {\bibinfo
  {volume} {109}},\ \bibinfo {pages} {193} (\bibinfo {year} {1998})},\ \Eprint
  {http://arxiv.org/abs/nucl-th/9709009} {arXiv:nucl-th/9709009 [nucl-th]}
  \BibitemShut {NoStop}%
\bibitem [{\citenamefont {Li}\ and\ \citenamefont {Ko}(1995)}]{Li:1995pra}%
  \BibitemOpen
  \bibfield  {author} {\bibinfo {author} {\bibfnamefont {B.-A.}\ \bibnamefont
  {Li}}\ and\ \bibinfo {author} {\bibfnamefont {C.~M.}\ \bibnamefont {Ko}},\
  }\href {\doibase 10.1103/PhysRevC.52.2037} {\bibfield  {journal} {\bibinfo
  {journal} {Phys. Rev.}\ }\textbf {\bibinfo {volume} {C52}},\ \bibinfo {pages}
  {2037} (\bibinfo {year} {1995})},\ \Eprint
  {http://arxiv.org/abs/nucl-th/9505016} {arXiv:nucl-th/9505016 [nucl-th]}
  \BibitemShut {NoStop}%
\bibitem [{\citenamefont {Wergieluk}\ \emph {et~al.}(2024)\citenamefont
  {Wergieluk}, \citenamefont {Weil}, \citenamefont {Tindall}, \citenamefont
  {Steinberg}, \citenamefont {Staudenmaier}, \citenamefont {Sorensen},
  \citenamefont {Sciarra}, \citenamefont {Schäfer}, \citenamefont {Sattler},
  \citenamefont {Ryu}, \citenamefont {Rothermel}, \citenamefont {Rose},
  \citenamefont {Roch}, \citenamefont {Prinz}, \citenamefont {Petersen},
  \citenamefont {Paulinyova}, \citenamefont {Pang}, \citenamefont
  {Oliinychenko}, \citenamefont {Mohs}, \citenamefont {Mitrovic}, \citenamefont
  {Mayer}, \citenamefont {Li}, \citenamefont {Kübler}, \citenamefont {Kretz},
  \citenamefont {Kehrenberg}, \citenamefont {Inghirami}, \citenamefont
  {Hirayama}, \citenamefont {Hammelmann}, \citenamefont {Götz}, \citenamefont
  {Groebel}, \citenamefont {Goldschmidt}, \citenamefont {Geiger}, \citenamefont
  {Garcia-Montero}, \citenamefont {Elfner}, \citenamefont {Ehlert},
  \citenamefont {Christensen}, \citenamefont {Bäuchle}, \citenamefont
  {Auvinen},\ and\ \citenamefont {Attems}}]{wergieluk_2024_10707746}%
  \BibitemOpen
  \bibfield  {author} {\bibinfo {author} {\bibfnamefont {A.}~\bibnamefont
  {Wergieluk}}, \bibinfo {author} {\bibfnamefont {J.}~\bibnamefont {Weil}},
  \bibinfo {author} {\bibfnamefont {J.}~\bibnamefont {Tindall}}, \bibinfo
  {author} {\bibfnamefont {V.}~\bibnamefont {Steinberg}}, \bibinfo {author}
  {\bibfnamefont {J.}~\bibnamefont {Staudenmaier}}, \bibinfo {author}
  {\bibfnamefont {A.}~\bibnamefont {Sorensen}}, \bibinfo {author}
  {\bibfnamefont {A.}~\bibnamefont {Sciarra}}, \bibinfo {author} {\bibfnamefont
  {A.}~\bibnamefont {Schäfer}}, \bibinfo {author} {\bibfnamefont
  {R.}~\bibnamefont {Sattler}}, \bibinfo {author} {\bibfnamefont
  {S.}~\bibnamefont {Ryu}}, \bibinfo {author} {\bibfnamefont {J.}~\bibnamefont
  {Rothermel}}, \bibinfo {author} {\bibfnamefont {J.-B.}\ \bibnamefont {Rose}},
  \bibinfo {author} {\bibfnamefont {H.}~\bibnamefont {Roch}}, \bibinfo {author}
  {\bibfnamefont {L.}~\bibnamefont {Prinz}}, \bibinfo {author} {\bibfnamefont
  {H.}~\bibnamefont {Petersen}}, \bibinfo {author} {\bibfnamefont
  {Z.}~\bibnamefont {Paulinyova}}, \bibinfo {author} {\bibfnamefont {L.-G.}\
  \bibnamefont {Pang}}, \bibinfo {author} {\bibfnamefont {D.}~\bibnamefont
  {Oliinychenko}}, \bibinfo {author} {\bibfnamefont {J.}~\bibnamefont {Mohs}},
  \bibinfo {author} {\bibfnamefont {D.}~\bibnamefont {Mitrovic}}, \bibinfo
  {author} {\bibfnamefont {M.}~\bibnamefont {Mayer}}, \bibinfo {author}
  {\bibfnamefont {F.}~\bibnamefont {Li}}, \bibinfo {author} {\bibfnamefont
  {N.}~\bibnamefont {Kübler}}, \bibinfo {author} {\bibfnamefont
  {M.}~\bibnamefont {Kretz}}, \bibinfo {author} {\bibfnamefont
  {T.}~\bibnamefont {Kehrenberg}}, \bibinfo {author} {\bibfnamefont
  {G.}~\bibnamefont {Inghirami}}, \bibinfo {author} {\bibfnamefont
  {R.}~\bibnamefont {Hirayama}}, \bibinfo {author} {\bibfnamefont
  {J.}~\bibnamefont {Hammelmann}}, \bibinfo {author} {\bibfnamefont
  {N.}~\bibnamefont {Götz}}, \bibinfo {author} {\bibfnamefont
  {J.}~\bibnamefont {Groebel}}, \bibinfo {author} {\bibfnamefont
  {A.}~\bibnamefont {Goldschmidt}}, \bibinfo {author} {\bibfnamefont
  {L.}~\bibnamefont {Geiger}}, \bibinfo {author} {\bibfnamefont
  {O.}~\bibnamefont {Garcia-Montero}}, \bibinfo {author} {\bibfnamefont
  {H.}~\bibnamefont {Elfner}}, \bibinfo {author} {\bibfnamefont
  {N.}~\bibnamefont {Ehlert}}, \bibinfo {author} {\bibfnamefont {C.~H.}\
  \bibnamefont {Christensen}}, \bibinfo {author} {\bibfnamefont
  {B.}~\bibnamefont {Bäuchle}}, \bibinfo {author} {\bibfnamefont
  {J.}~\bibnamefont {Auvinen}}, \ and\ \bibinfo {author} {\bibfnamefont
  {M.}~\bibnamefont {Attems}},\ }\href {\doibase 10.5281/zenodo.10707746}
  {\enquote {\bibinfo {title} {smash-transport/smash: Smash-3.1},}\ } (\bibinfo
  {year} {2024})\BibitemShut {NoStop}%
\bibitem [{\citenamefont {Bass}\ \emph {et~al.}(1998)\citenamefont {Bass} \emph
  {et~al.}}]{Bass:1998ca}%
  \BibitemOpen
  \bibfield  {author} {\bibinfo {author} {\bibfnamefont {S.~A.}\ \bibnamefont
  {Bass}} \emph {et~al.},\ }\href {\doibase 10.1016/S0146-6410(98)00058-1}
  {\bibfield  {journal} {\bibinfo  {journal} {Prog. Part. Nucl. Phys.}\
  }\textbf {\bibinfo {volume} {41}},\ \bibinfo {pages} {255} (\bibinfo {year}
  {1998})},\ \Eprint {http://arxiv.org/abs/nucl-th/9803035}
  {arXiv:nucl-th/9803035} \BibitemShut {NoStop}%
\bibitem [{\citenamefont {Sj\"ostrand}\ \emph {et~al.}(2015)\citenamefont
  {Sj\"ostrand}, \citenamefont {Ask}, \citenamefont {Christiansen},
  \citenamefont {Corke}, \citenamefont {Desai}, \citenamefont {Ilten},
  \citenamefont {Mrenna}, \citenamefont {Prestel}, \citenamefont {Rasmussen},\
  and\ \citenamefont {Skands}}]{Sjostrand:2014zea}%
  \BibitemOpen
  \bibfield  {author} {\bibinfo {author} {\bibfnamefont {T.}~\bibnamefont
  {Sj\"ostrand}}, \bibinfo {author} {\bibfnamefont {S.}~\bibnamefont {Ask}},
  \bibinfo {author} {\bibfnamefont {J.~R.}\ \bibnamefont {Christiansen}},
  \bibinfo {author} {\bibfnamefont {R.}~\bibnamefont {Corke}}, \bibinfo
  {author} {\bibfnamefont {N.}~\bibnamefont {Desai}}, \bibinfo {author}
  {\bibfnamefont {P.}~\bibnamefont {Ilten}}, \bibinfo {author} {\bibfnamefont
  {S.}~\bibnamefont {Mrenna}}, \bibinfo {author} {\bibfnamefont
  {S.}~\bibnamefont {Prestel}}, \bibinfo {author} {\bibfnamefont {C.~O.}\
  \bibnamefont {Rasmussen}}, \ and\ \bibinfo {author} {\bibfnamefont {P.~Z.}\
  \bibnamefont {Skands}},\ }\href {\doibase 10.1016/j.cpc.2015.01.024}
  {\bibfield  {journal} {\bibinfo  {journal} {Comput. Phys. Commun.}\ }\textbf
  {\bibinfo {volume} {191}},\ \bibinfo {pages} {159} (\bibinfo {year}
  {2015})},\ \Eprint {http://arxiv.org/abs/1410.3012} {arXiv:1410.3012
  [hep-ph]} \BibitemShut {NoStop}%
\bibitem [{\citenamefont {Bellm}\ \emph {et~al.}(2016)\citenamefont {Bellm}
  \emph {et~al.}}]{Bellm:2015jjp}%
  \BibitemOpen
  \bibfield  {author} {\bibinfo {author} {\bibfnamefont {J.}~\bibnamefont
  {Bellm}} \emph {et~al.},\ }\href {\doibase 10.1140/epjc/s10052-016-4018-8}
  {\bibfield  {journal} {\bibinfo  {journal} {Eur. Phys. J. C}\ }\textbf
  {\bibinfo {volume} {76}},\ \bibinfo {pages} {196} (\bibinfo {year} {2016})},\
  \Eprint {http://arxiv.org/abs/1512.01178} {arXiv:1512.01178 [hep-ph]}
  \BibitemShut {NoStop}%
\bibitem [{\citenamefont {Hoeche}\ \emph {et~al.}(2013)\citenamefont {Hoeche},
  \citenamefont {Krauss}, \citenamefont {Schonherr},\ and\ \citenamefont
  {Siegert}}]{Hoeche:2012yf}%
  \BibitemOpen
  \bibfield  {author} {\bibinfo {author} {\bibfnamefont {S.}~\bibnamefont
  {Hoeche}}, \bibinfo {author} {\bibfnamefont {F.}~\bibnamefont {Krauss}},
  \bibinfo {author} {\bibfnamefont {M.}~\bibnamefont {Schonherr}}, \ and\
  \bibinfo {author} {\bibfnamefont {F.}~\bibnamefont {Siegert}},\ }\href
  {\doibase 10.1007/JHEP04(2013)027} {\bibfield  {journal} {\bibinfo  {journal}
  {JHEP}\ }\textbf {\bibinfo {volume} {04}},\ \bibinfo {pages} {027} (\bibinfo
  {year} {2013})},\ \Eprint {http://arxiv.org/abs/1207.5030} {arXiv:1207.5030
  [hep-ph]} \BibitemShut {NoStop}%
\bibitem [{\citenamefont {B\'\i{}r\'o}\ \emph {et~al.}(2019)\citenamefont
  {B\'\i{}r\'o}, \citenamefont {Barnaf\"oldi}, \citenamefont {Papp},
  \citenamefont {Gyulassy}, \citenamefont {L\'evai}, \citenamefont {Wang},\
  and\ \citenamefont {Zhang}}]{Biro:2019ijx}%
  \BibitemOpen
  \bibfield  {author} {\bibinfo {author} {\bibfnamefont {G.}~\bibnamefont
  {B\'\i{}r\'o}}, \bibinfo {author} {\bibfnamefont {G.~G.}\ \bibnamefont
  {Barnaf\"oldi}}, \bibinfo {author} {\bibfnamefont {G.}~\bibnamefont {Papp}},
  \bibinfo {author} {\bibfnamefont {M.}~\bibnamefont {Gyulassy}}, \bibinfo
  {author} {\bibfnamefont {P.}~\bibnamefont {L\'evai}}, \bibinfo {author}
  {\bibfnamefont {X.-N.}\ \bibnamefont {Wang}}, \ and\ \bibinfo {author}
  {\bibfnamefont {B.-W.}\ \bibnamefont {Zhang}},\ }\href {\doibase
  10.22323/1.345.0045} {\bibfield  {journal} {\bibinfo  {journal} {PoS}\
  }\textbf {\bibinfo {volume} {HardProbes2018}},\ \bibinfo {pages} {045}
  (\bibinfo {year} {2019})},\ \Eprint {http://arxiv.org/abs/1901.04220}
  {arXiv:1901.04220 [physics.comp-ph]} \BibitemShut {NoStop}%
\bibitem [{\citenamefont {Krauss}\ \emph {et~al.}(2019)\citenamefont {Krauss},
  \citenamefont {Lindert}, \citenamefont {Linten},\ and\ \citenamefont
  {Sch\"onherr}}]{Krauss:2018djz}%
  \BibitemOpen
  \bibfield  {author} {\bibinfo {author} {\bibfnamefont {F.}~\bibnamefont
  {Krauss}}, \bibinfo {author} {\bibfnamefont {J.~M.}\ \bibnamefont {Lindert}},
  \bibinfo {author} {\bibfnamefont {R.}~\bibnamefont {Linten}}, \ and\ \bibinfo
  {author} {\bibfnamefont {M.}~\bibnamefont {Sch\"onherr}},\ }\href {\doibase
  10.1140/epjc/s10052-019-6614-x} {\bibfield  {journal} {\bibinfo  {journal}
  {Eur. Phys. J. C}\ }\textbf {\bibinfo {volume} {79}},\ \bibinfo {pages} {143}
  (\bibinfo {year} {2019})},\ \Eprint {http://arxiv.org/abs/1809.10650}
  {arXiv:1809.10650 [hep-ph]} \BibitemShut {NoStop}%
\bibitem [{\citenamefont {H\"oche}\ \emph {et~al.}(2018)\citenamefont
  {H\"oche}, \citenamefont {Kuttimalai},\ and\ \citenamefont
  {Li}}]{Hoche:2018gti}%
  \BibitemOpen
  \bibfield  {author} {\bibinfo {author} {\bibfnamefont {S.}~\bibnamefont
  {H\"oche}}, \bibinfo {author} {\bibfnamefont {S.}~\bibnamefont {Kuttimalai}},
  \ and\ \bibinfo {author} {\bibfnamefont {Y.}~\bibnamefont {Li}},\ }\href
  {\doibase 10.1103/PhysRevD.98.114013} {\bibfield  {journal} {\bibinfo
  {journal} {Phys. Rev. D}\ }\textbf {\bibinfo {volume} {98}},\ \bibinfo
  {pages} {114013} (\bibinfo {year} {2018})},\ \Eprint
  {http://arxiv.org/abs/1809.04192} {arXiv:1809.04192 [hep-ph]} \BibitemShut
  {NoStop}%
\bibitem [{\citenamefont {Hoeche}\ \emph {et~al.}(2012)\citenamefont {Hoeche},
  \citenamefont {Krauss}, \citenamefont {Schonherr},\ and\ \citenamefont
  {Siegert}}]{Hoeche:2011fd}%
  \BibitemOpen
  \bibfield  {author} {\bibinfo {author} {\bibfnamefont {S.}~\bibnamefont
  {Hoeche}}, \bibinfo {author} {\bibfnamefont {F.}~\bibnamefont {Krauss}},
  \bibinfo {author} {\bibfnamefont {M.}~\bibnamefont {Schonherr}}, \ and\
  \bibinfo {author} {\bibfnamefont {F.}~\bibnamefont {Siegert}},\ }\href
  {\doibase 10.1007/JHEP09(2012)049} {\bibfield  {journal} {\bibinfo  {journal}
  {JHEP}\ }\textbf {\bibinfo {volume} {09}},\ \bibinfo {pages} {049} (\bibinfo
  {year} {2012})},\ \Eprint {http://arxiv.org/abs/1111.1220} {arXiv:1111.1220
  [hep-ph]} \BibitemShut {NoStop}%
\bibitem [{\citenamefont {Buckley}\ and\ \citenamefont
  {Bakshi~Gupta}(2016)}]{Buckley:2016bhy}%
  \BibitemOpen
  \bibfield  {author} {\bibinfo {author} {\bibfnamefont {A.}~\bibnamefont
  {Buckley}}\ and\ \bibinfo {author} {\bibfnamefont {D.}~\bibnamefont
  {Bakshi~Gupta}},\ }\href@noop {} {\  (\bibinfo {year} {2016})},\ \Eprint
  {http://arxiv.org/abs/1608.03577} {arXiv:1608.03577 [hep-ph]} \BibitemShut
  {NoStop}%
\bibitem [{\citenamefont {Neill}\ \emph {et~al.}(2019)\citenamefont {Neill},
  \citenamefont {Papaefstathiou}, \citenamefont {Waalewijn},\ and\
  \citenamefont {Zoppi}}]{Neill:2018wtk}%
  \BibitemOpen
  \bibfield  {author} {\bibinfo {author} {\bibfnamefont {D.}~\bibnamefont
  {Neill}}, \bibinfo {author} {\bibfnamefont {A.}~\bibnamefont
  {Papaefstathiou}}, \bibinfo {author} {\bibfnamefont {W.~J.}\ \bibnamefont
  {Waalewijn}}, \ and\ \bibinfo {author} {\bibfnamefont {L.}~\bibnamefont
  {Zoppi}},\ }\href {\doibase 10.1007/JHEP01(2019)067} {\bibfield  {journal}
  {\bibinfo  {journal} {JHEP}\ }\textbf {\bibinfo {volume} {01}},\ \bibinfo
  {pages} {067} (\bibinfo {year} {2019})},\ \Eprint
  {http://arxiv.org/abs/1810.12915} {arXiv:1810.12915 [hep-ph]} \BibitemShut
  {NoStop}%
\bibitem [{\citenamefont {Reyer}\ \emph {et~al.}(2019)\citenamefont {Reyer},
  \citenamefont {Sch\"onherr},\ and\ \citenamefont {Schumann}}]{Reyer:2019obz}%
  \BibitemOpen
  \bibfield  {author} {\bibinfo {author} {\bibfnamefont {M.}~\bibnamefont
  {Reyer}}, \bibinfo {author} {\bibfnamefont {M.}~\bibnamefont {Sch\"onherr}},
  \ and\ \bibinfo {author} {\bibfnamefont {S.}~\bibnamefont {Schumann}},\
  }\href {\doibase 10.1140/epjc/s10052-019-6815-3} {\bibfield  {journal}
  {\bibinfo  {journal} {Eur. Phys. J. C}\ }\textbf {\bibinfo {volume} {79}},\
  \bibinfo {pages} {321} (\bibinfo {year} {2019})},\ \Eprint
  {http://arxiv.org/abs/1902.01763} {arXiv:1902.01763 [hep-ph]} \BibitemShut
  {NoStop}%
\bibitem [{\citenamefont {Bothmann}\ and\ \citenamefont
  {Debbio}(2019)}]{Bothmann:2018trh}%
  \BibitemOpen
  \bibfield  {author} {\bibinfo {author} {\bibfnamefont {E.}~\bibnamefont
  {Bothmann}}\ and\ \bibinfo {author} {\bibfnamefont {L.}~\bibnamefont
  {Debbio}},\ }\href {\doibase 10.1007/JHEP01(2019)033} {\bibfield  {journal}
  {\bibinfo  {journal} {JHEP}\ }\textbf {\bibinfo {volume} {01}},\ \bibinfo
  {pages} {033} (\bibinfo {year} {2019})},\ \Eprint
  {http://arxiv.org/abs/1808.07802} {arXiv:1808.07802 [hep-ph]} \BibitemShut
  {NoStop}%
\bibitem [{Pro(2018)}]{Proceedings:2018jsb}%
  \BibitemOpen
  \href@noop {} {\emph {\bibinfo {title} {{Les Houches 2017: Physics at TeV
  Colliders Standard Model Working Group Report}}}}\ (\bibinfo {year} {2018})\
  \Eprint {http://arxiv.org/abs/1803.07977} {arXiv:1803.07977 [hep-ph]}
  \BibitemShut {NoStop}%
\bibitem [{\citenamefont {de~Florian}\ \emph {et~al.}(2016)\citenamefont
  {de~Florian} \emph {et~al.}}]{LHCHiggsCrossSectionWorkingGroup:2016ypw}%
  \BibitemOpen
  \bibfield  {author} {\bibinfo {author} {\bibfnamefont {D.}~\bibnamefont
  {de~Florian}} \emph {et~al.} (\bibinfo {collaboration} {LHC Higgs Cross
  Section Working Group}),\ }\href {\doibase 10.23731/CYRM-2017-002} {\ \textbf
  {\bibinfo {volume} {2/2017}} (\bibinfo {year} {2016}),\
  10.23731/CYRM-2017-002},\ \Eprint {http://arxiv.org/abs/1610.07922}
  {arXiv:1610.07922 [hep-ph]} \BibitemShut {NoStop}%
\bibitem [{\citenamefont {Dobbs}\ and\ \citenamefont
  {Hansen}(2001)}]{Dobbs:2001ck}%
  \BibitemOpen
  \bibfield  {author} {\bibinfo {author} {\bibfnamefont {M.}~\bibnamefont
  {Dobbs}}\ and\ \bibinfo {author} {\bibfnamefont {J.~B.}\ \bibnamefont
  {Hansen}},\ }\href {\doibase 10.1016/S0010-4655(00)00189-2} {\bibfield
  {journal} {\bibinfo  {journal} {Comput. Phys. Commun.}\ }\textbf {\bibinfo
  {volume} {134}},\ \bibinfo {pages} {41} (\bibinfo {year} {2001})}\BibitemShut
  {NoStop}%
\bibitem [{\citenamefont {Buckley}\ \emph {et~al.}(2021)\citenamefont
  {Buckley}, \citenamefont {Ilten}, \citenamefont {Konstantinov}, \citenamefont
  {L\"onnblad}, \citenamefont {Monk}, \citenamefont {Pokorski}, \citenamefont
  {Przedzinski},\ and\ \citenamefont {Verbytskyi}}]{Buckley:2019xhk}%
  \BibitemOpen
  \bibfield  {author} {\bibinfo {author} {\bibfnamefont {A.}~\bibnamefont
  {Buckley}}, \bibinfo {author} {\bibfnamefont {P.}~\bibnamefont {Ilten}},
  \bibinfo {author} {\bibfnamefont {D.}~\bibnamefont {Konstantinov}}, \bibinfo
  {author} {\bibfnamefont {L.}~\bibnamefont {L\"onnblad}}, \bibinfo {author}
  {\bibfnamefont {J.}~\bibnamefont {Monk}}, \bibinfo {author} {\bibfnamefont
  {W.}~\bibnamefont {Pokorski}}, \bibinfo {author} {\bibfnamefont
  {T.}~\bibnamefont {Przedzinski}}, \ and\ \bibinfo {author} {\bibfnamefont
  {A.}~\bibnamefont {Verbytskyi}},\ }\href {\doibase 10.1016/j.cpc.2020.107310}
  {\bibfield  {journal} {\bibinfo  {journal} {Comput. Phys. Commun.}\ }\textbf
  {\bibinfo {volume} {260}},\ \bibinfo {pages} {107310} (\bibinfo {year}
  {2021})},\ \Eprint {http://arxiv.org/abs/1912.08005} {arXiv:1912.08005
  [hep-ph]} \BibitemShut {NoStop}%
\bibitem [{\citenamefont {Bierlich}\ \emph
  {et~al.}(2020{\natexlab{b}})\citenamefont {Bierlich} \emph
  {et~al.}}]{Bierlich:2020wms}%
  \BibitemOpen
  \bibfield  {author} {\bibinfo {author} {\bibfnamefont {C.}~\bibnamefont
  {Bierlich}} \emph {et~al.},\ }\href {\doibase 10.1140/epjc/s10052-020-8033-4}
  {\bibfield  {journal} {\bibinfo  {journal} {Eur. Phys. J. C}\ }\textbf
  {\bibinfo {volume} {80}},\ \bibinfo {pages} {485} (\bibinfo {year}
  {2020}{\natexlab{b}})},\ \Eprint {http://arxiv.org/abs/2001.10737}
  {arXiv:2001.10737 [hep-ph]} \BibitemShut {NoStop}%
\bibitem [{\citenamefont {Mohs}\ \emph {et~al.}(2020)\citenamefont {Mohs},
  \citenamefont {Ryu},\ and\ \citenamefont {Elfner}}]{Mohs:2019iee}%
  \BibitemOpen
  \bibfield  {author} {\bibinfo {author} {\bibfnamefont {J.}~\bibnamefont
  {Mohs}}, \bibinfo {author} {\bibfnamefont {S.}~\bibnamefont {Ryu}}, \ and\
  \bibinfo {author} {\bibfnamefont {H.}~\bibnamefont {Elfner}} (\bibinfo
  {collaboration} {SMASH}),\ }\href {\doibase 10.1088/1361-6471/ab7bd1}
  {\bibfield  {journal} {\bibinfo  {journal} {J. Phys. G}\ }\textbf {\bibinfo
  {volume} {47}},\ \bibinfo {pages} {065101} (\bibinfo {year} {2020})},\
  \Eprint {http://arxiv.org/abs/1909.05586} {arXiv:1909.05586 [nucl-th]}
  \BibitemShut {NoStop}%
\end{thebibliography}%
\end{document}